\documentclass{article}

\usepackage{arxiv}

\usepackage[utf8]{inputenc} 
\usepackage[T1]{fontenc}    
\usepackage{hyperref}       
\usepackage{url}            
\usepackage{booktabs}       
\usepackage{amsfonts}       
\usepackage{nicefrac}       
\usepackage{microtype}      
\usepackage{lipsum}		
\usepackage{graphicx}
\usepackage{doi}
\usepackage{amsmath}
\usepackage{algorithm}
\usepackage[noend]{algpseudocode}
\usepackage{caption}
\usepackage{subcaption}
\usepackage{tikz}
\usepackage{enumitem}
\usepackage[normalem]{ulem}
\newcommand*\circled[1]{\tikz[baseline=(char.base)]{
            \node[shape=circle,draw,inner sep=1pt] (char) {#1};}}
\newcommand{\fullsys}{Adaptive Hotspot-Aware Tree}
\newcommand{\sys}{AHA-tree}
\newcommand{\sysw}{AHA-W$^0$}
\newcommand{\syswp}{AHA-W$^+$}
\newcommand{\sysr}{AHA-R}

\newcommand{\bplustree}{B$^{+}$-tree}
\newcommand{\lsmt}{LSM-tree}
\newcommand{\sst}{SSTable}

\newcommand{\rtl}{\texttt{rootLSM-tree}}
\newcommand{\ndl}{\texttt{nodeLSM-tree}}

\newcommand{\lsmtcomp}{LSMT-component}

\newcommand{\memtab}{\texttt{MemTable}}
\title{An Adaptive Hotspot-Aware Index for Oscillating Write-Heavy and Read-Heavy Workloads}

\author{Lu Xing\\
	Department of Computer Science\\
	Purdue University\\
	West Lafayette, IN 47906 \\
	\texttt{xingl@purdue.edu} \\
	\And
	{Ruihong Wang}\\
	Department of Computer Science\\
	Purdue University\\
	West Lafayette, IN 47906 \\
	\texttt{wang4996@purdue.edu} \\
	\AND
        Walid G. Aref\\
	Department of Computer Science\\
	Purdue University\\
	West Lafayette, IN 47906 \\
	\texttt{aref@purdue.edu} \\
}

\date{}

\begin{document}
\maketitle

\begin{abstract}
HTAP systems are designed to handle transactional and analytical workloads. Besides a mixed workload at any given time, the workload can also change over time. A popular type of continuously changing workload is one that oscillates between being write-heavy at times and being read-heavy at other times. Oscillating workloads can be observed in many applications. Indexes, e.g., the \bplustree{} and the \lsmt{}, cannot perform equally well all the time. Conventional adaptive indexing does not solve this issue as it focuses on adapting in one direction. This paper studies how to support oscillating workloads with adaptive indexes that adapt the underlying index structures in both directions. With the observation that real-world datasets are skewed, the focus is to optimize the index within the hotspot regions. 
The \fullsys{} (or \sys{}, for short) is introduced, where its
adaptation is bi-directional.
Experimental evaluation 
show that \sys{} can behave competitively as compared to an \lsmt{} for write-heavy transactional workloads. Upon switching to a read-heavy analytical workload, \sys{} can gradually adapt and behave competitively,
and can match the \bplustree{}'s read performance. 
\end{abstract}


\section{Introduction}\label{section:introduction}

Nowadays, database management systems are not just built for a single purpose, but rather they face various requirements from users. Hybrid Transactional and Analytical Processing systems (HTAP, for short)  are becoming more popular as they address hybrid requirements. These hybrid requirements include transactional processing as well as analytical queries. Even in a data warehouse scenario, a heterogeneous workload, including bulk-loading, periodical data refreshing, and complex queries, is expected~\cite{chaudhuri1997overview, dageville2016snowflake}. However, the heterogeneous workload is not static 
over time.

One of the examples of 
workloads that change
over time is the TPC-DS benchmark~\cite{tpcds}, the well-known decision support benchmark.
TPC-DS includes three different disciplines that are executed in order; database load, query run, and data maintenance followed by another query run~\cite{poess2007why}. Data maintenance includes inserts and deletes~\cite{nambiar2006the}. Another example is the HTAP system benchmark~\cite{zhanghybench} that includes workloads  featuring transactions first, then analytical queries next. 
Agrawal et al.~\cite{agrawal2006sigmod} describe
data warehousing as ``query by day, update at night", and  oscillates between these two modes of operation. The same notion is  mentioned in C-store~\cite{Stonebraker2005cstore}, where a data warehouse periodically performs a bulk load of new data followed by a relatively long period of ad-hoc analytical queries. The change in workload is also observed as a diurnal pattern in one of the Rocksdb use cases at Meta~\cite{cao2020characterizing}, and is also observed in~\cite{daghistani2021swarm}, where the number of tweets fluctuate across the day. Workloads with daily patterns are also observed in~\cite{atikoglu2012workload,curino2011workload,gmach2007workload,ma2018query}.

One category of the changing workloads is the oscillating workload that is write-heavy at times and is read-heavy at other times. Agrawal et al. present a data warehousing scenario in Figure 1 of~\cite{agrawal2006sigmod} 
that complex queries are issued during the day and the warehouse data is updated in a batch window at night. In social media applications, users post and comment actively during the day, and browse content in the night or early in the morning. This corresponds to the write-heavy (post and comment) and read-heavy (browse) workloads that keep repeating every day. Similar examples can be found in the context of traffic incidents management. When incidents happen, there can be a surge in the write operations while at other times a read-heavy workload is the dominant one.

Real-world datasets 
contain hotspots, where data is frequently accessed, e.g., popular figures on social media, and navigating maps in downtown area. \cite{cao2020characterizing} uncovers that hot data is usually located closely in a small range. And only a small portion of the data is hot. Moreover, there can be multiple hotspots within the dataset~\cite{cao2020characterizing}. Besides, popularity of hot data may evolve over time~\cite{zhang2022sa}.

\noindent
\textbf{Challenges.}
Traditional non-adaptive indexes, e.g., 
the Log-Structured Merge Tree (\lsmt{})~\cite{o1996log} is optimized for only one operation, i.e., 
the
write operation. 
The \bplustree{}~\cite{bayer1970organization,comer1979ubiquitous} has good read performance, but is not write-optimized when compared to the \lsmt{}.
Neither index can 
perform
equally well in  oscillating write-heavy and read-heavy workloads.

One way to handle  hotspots is to have two separate indexes, one dedicated for the read hotspot (frequently read but less frequently updated), and one for the remaining cold data (less frequently read). The hotspot index can be designed to be read-optimized for the read hotspot and the cold data index be write-optimized as reading is not that frequent. 
However, when the hotspot changes, this method suffers because re-constructing indexes incurs much overhead. Also, when there are multiple hotspots, keeping track of data ranges and dispatching data require a routing layer that adds more overhead.

\begin{figure}[h]
    \centering
    \begin{subfigure}{0.49\linewidth}
    \centering
        \includegraphics[width=\linewidth]{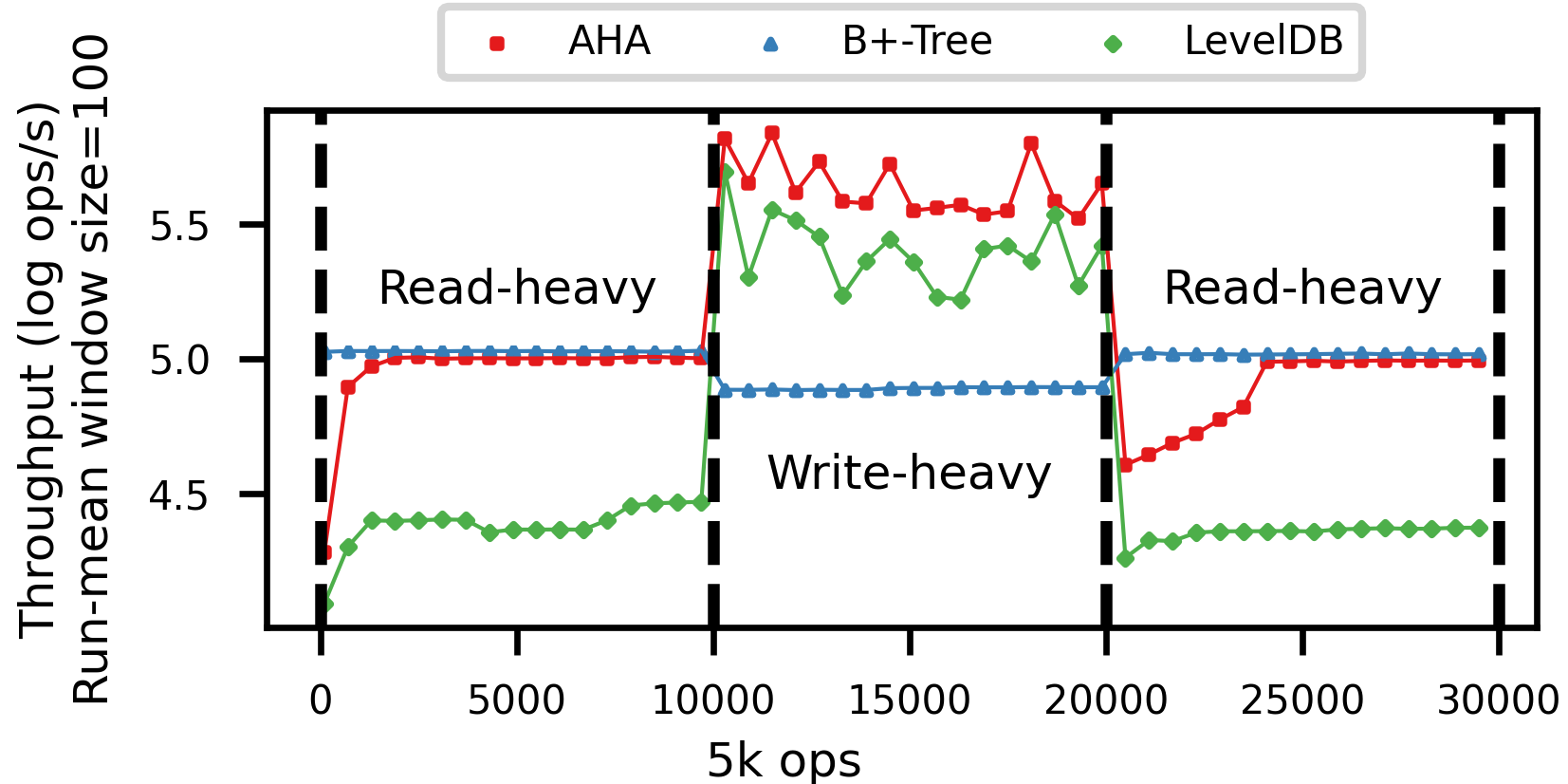}
        \caption{Oscillating workload}
        \label{fig:teaser_a}
    \end{subfigure}
    \hfill
    \begin{subfigure}{0.49\linewidth}
    \centering
        \includegraphics[width=\linewidth]{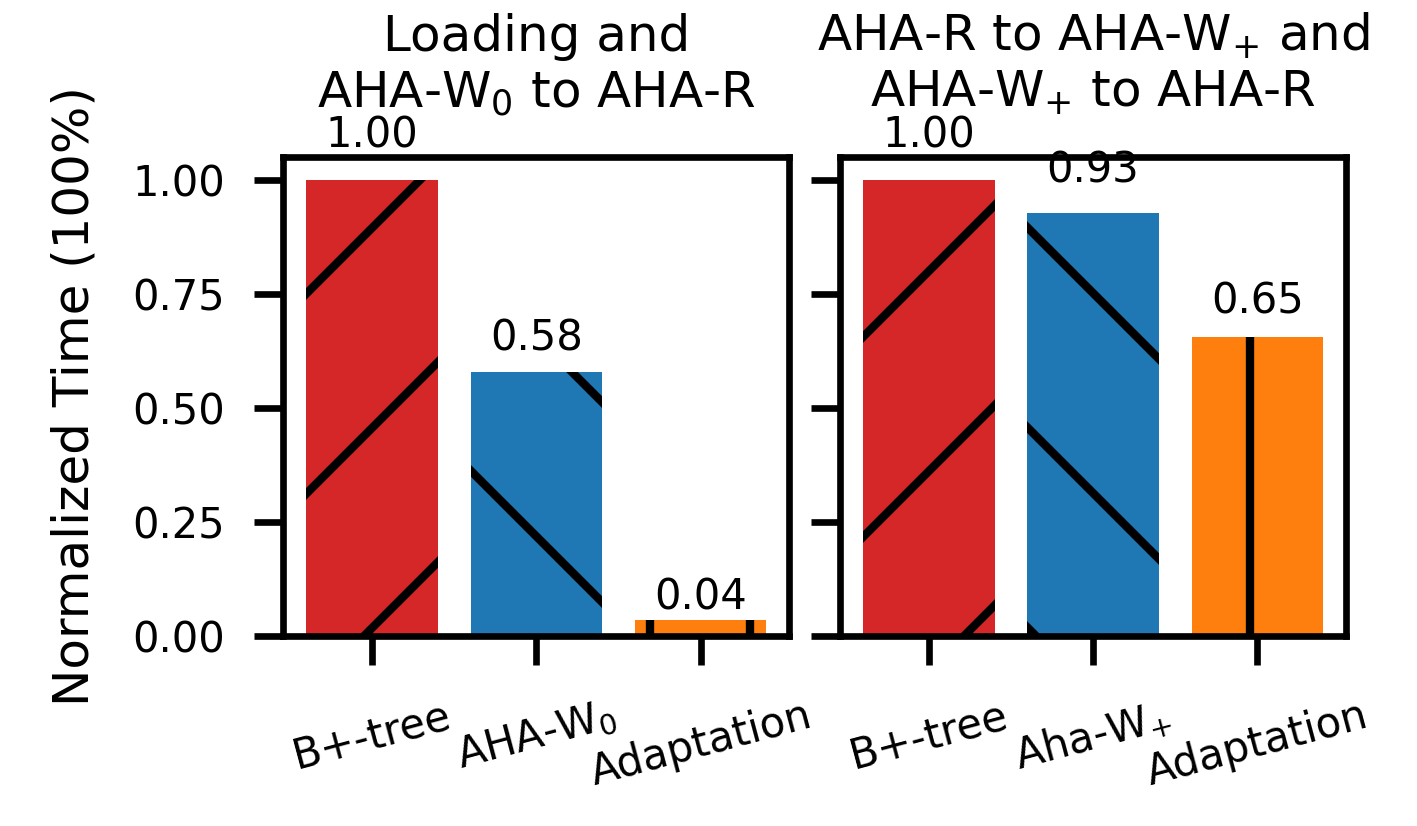}
        \caption{Relative time of \sys{}}
        \label{fig:teaser_b}
    \end{subfigure}
    \caption{Performance of \sys{} (with optimization mentioned in Section~\ref{sssection:potential_opt}) and baselines under oscillating write- and read-heavy workloads.}
    \label{fig:teaser}
\end{figure}

\noindent
\textbf{The \sys{} Approach.} 
It would be ideal  for an oscillating workload to have an  index that
behaves 
as an \lsmt{} during the write-heavy phase, 
and behave as a 
\bplustree{} during the read-heavy workload. As 
in Figure~\ref{fig:teaser_a}, \sys{} reads like 
a \bplustree{} (during Operations 0-50 Million and 100-150 Million) and writes like an \lsmt{} (during Operations 50-100 Million). \sys{} is inspired by the 
buffer tree~\cite{arge2003buffer}, where it 
has
a tree part (a \bplustree{}) 
and a
buffer part (an \lsmt{}). 
The buffer tree~\cite{arge2003buffer} may not solve the problem of oscillating workloads as it is 
a non-adaptive structure. To make the buffer tree adaptive, we let the index adapt itself in either workload. During the read-heavy workload, hotspot data is reorganized to be sequentially sorted, such that range searches probe the index like a \bplustree{}. During the write-heavy workload, data is inserted in batches to avoid I/Os caused by individual inserts.

The \sys{} is a new adaptive index  introduced in this paper. 
We test its adaptation under  oscillating write- and read-heavy workloads. The \sys{} is a hierarchical structure that has two components: an \lsmtcomp{} and a tree structure of tree pages. The \lsmtcomp{} is divided into multiple mini-components,
each 
associated with a tree page. \sys{} is hotspot-aware, i.e., 
makes the adaptation process confined to the hotspot region. 

The contributions of this paper 
are 
as follows:
\begin{itemize}
    \item We introduce a bi-directional adaptive index, the \sys{}, that is suited for workloads oscillating between being read-heavy at times and being write-heavy at other times. 
    \item The \sys{} handles concurrent read and write operations as well as adapts to workload changes with concurrent background threads.
    \item The \sys{} can also adapt itself under diverse hotspot situations, including multiple hotspots, and when the hotspots change their location.
\end{itemize}

The rest of this paper proceeds as follows. Section~\ref{section:related} discusses  related work. Section~\ref{section:motivation} introduces the motivation and background for adaptive indexing. An overview of \sys{} 
is presented in Section~\ref{section:overview}.  
The various states and modes of operations of the \sys{} 
are described in Sections~\ref{section:w0}, \ref{section:r}, and~\ref{section:wp}. Section~\ref{section:implementation} presents the implementation and optimization of \sys{}. Section~\ref{section:transition} describes the bidirectional transitions among the write-optimized  and read-optimized states of the \sys{}. 
Section~\ref{section:experiment} presents the experimental results, 
and, Section~\ref{section:conclusion} concludes the paper.


\section{Related Work}\label{section:related}
\noindent
{\bf Write-optimized Indexes}. An enormous number of studies have been conducted to improve the write performance of tree-like structures. To reduce I/O, the buffer tree~\cite{arge2003buffer} batches multiple operations (inserts, deletes, and range searches) into one segment in  memory. The design of~\cite{graefe2004write} makes page migration inexpensive in log-structured systems, and supports both in-place updates and large append-only writes at the same time. B$\epsilon$-tree~\cite{bender2015introduction} is an in-memory tree structure, where some space of the tree node is allocated to buffer operations. 
Zeighami~\cite{zeighami2019nested} presents a Nested B-tree, where each B-tree node contains a \bplustree{}. Bf-tree~\cite{hao2024bf} is read-write-optimized by building a variable length buffer pool. There are also hardware-specific write-optimized tree structures~\cite{luo2021tlb,wang2022sherman,conway2020splinterdb,wang2021wobtree,an2023marlin}.

\noindent
{\bf Range Query-optimized \lsmt{}}. Since the \lsmt{} sacrifices read performance in exchange for better write performance, research has been conducted to improve its read performance. A bloom filter~\cite{bloom1970space} can only improve point reads of the \lsmt{} by filtering and reducing unnecessary I/Os. Rosetta~\cite{luo2020rosetta} uses a hierarchically-stacked bloom filters as a range filter. REMIX~\cite{zhong2021remix} improves range searches by adding a sorted view across multiple \sst{}s in the \lsmt{} s.t. range search can find the target key using binary search, and retrieve following keys in order without comparison.

\noindent
{\bf Database Cracking}. In database cracking~\cite{idreos2007database}, every query helps partially sort the column in a column-store. To support dynamic databases, Idreas et al.~\cite{idreos2007updating} propose several algorithms to update a cracked database. Partial sideways cracking~\cite{idreos2009self} builds on top of data cracking, and minimizes tuple reconstruction cost. As database cracking is sensitive to the past workload, stochastic cracking~\cite{halim2012stochastic} is proposed to increase its robustness. 
As the query performance of database cracking may degrade when a less refined part of the index is queried, progressive indexing solves the above by controlling the indexing budget and offers predictable performance~\cite{holanda2019progressive}.

\noindent
{\bf Other Adaptive Indexes}.
The splay tree~\cite{sleator1985splay} can move the recently accessed data to the root. 
Idreos et al.~\cite{idreos2019design} show a transformation between the \lsmt{} and the \bplustree{} but requires the \lsmt{} to store all its data only in the bottom level, and a \bplustree{} can be bulk loaded out of it. Adaptive Hybrid Indexes~\cite{anneser2022adaptive} are introduced to address the memory overhead of compact indexes to deal with changes in data access. VIP-hashing~\cite{kakaraparthy2022vip} deals with the data hotness issue in the hash table by learning the data distribution as more queries arrive. SA-LSM~\cite{zhang2022sa} uses survival analysis to train a model to predict the next access time of the data item, and moves cold data into slower storage (HDD compared to Enhanced SSD) during \lsmt{} compaction. A real-time \lsmt{}, termed LASER~\cite{saxena2023real}, is used for a given HTAP workload by designing an optimal mixed data layout of the \lsmt{}. B$^{link}$-hash~\cite{cha2023blink} solves the issue of inserting in-memory monotonically increasing keys in the \bplustree{} by using a hash table as a leaf node in the tree that later adapts to a leaf page upon querying. 
Mo et al.~\cite{mo2023flsm} optimizes the \lsmt{} within its design space via reinforcement learning. SALI~\cite{ge2023sali} is a scalable adaptive learned index framework that can adapt its nodes to the workload.

Among the above, \cite{sleator1985splay,anneser2022adaptive,kakaraparthy2022vip} focus on being adaptive to the data access frequency. \cite{idreos2007database,halim2012stochastic,cha2023blink} focus on being adaptive in one direction. \cite{saxena2023real, chatterjee2021cosine, chatterjee2024limo} focus on finding the optimal index design given a workload. \cite{mo2023flsm, zhang2022sa} focus on being adaptive within the design space of a single index type. Most of them~\cite{ge2023sali, kakaraparthy2022vip,cha2023blink, anneser2022adaptive} focus on in-memory adaptive indexing. None of the existing techniques can be applied directly to  oscillating workloads.

\section{Motivation}\label{section:motivation}

An oscillating read- and write-heavy workload is  typical  for HTAP systems. Write-heavy operations involve fast ingestion of data while read-heavy operations involve analysis of the newly ingested data. This workload pattern is challenging for  non-adaptive indexes including the \bplustree{} and the \lsmt{}. 

In Figure~\ref{fig:teaser}, we load all the indexes with the same data, then issue 50 Million range search queries over the hotspot, followed by 50 Million write operations, then 50 Million range search queries over the hotspot. The experiment's details are  in Section~\ref{section:experiment}. During Operations 0-50 Million, the \bplustree{} shows the best range search throughput, while the \lsmt{} performs the worst in Figure~\ref{fig:teaser_a}. \sys{} gradually adapts itself, achieving eventual throughput nearly equal to that of \bplustree{} and 3.4$\times$ that of \lsmt{}. During Operations 50-100 Million, each index is written with the same amount of data. This time, the average throughput of the \lsmt{} is the best (Figure~\ref{fig:teaser_a}). During Operations 100-150 Million, the workload oscillates to range search queries in the hotspot, the performance trend of all compared indexes are the same as in the first range search phase, though \sys{} exhibits a more noticeable transition period.

Figure~\ref{fig:teaser_b} shows that the relative time of \sys{} index construction is 58\% of \bplustree{} construction time; finishing 50 Million write operations in \sys{} uses 93\% of \bplustree{} time. The first adaptation time is 4\% of \bplustree{} construction time while the second adaptation time is 65\% of \bplustree{} write-phase time.

From Figure~\ref{fig:teaser}, it is almost impossible for an index to perform equally well in both read- and write-heavy workloads, given that the index is optimized for either write operation or read operation. For example, the \lsmt{} performs well during the write-heavy workload but poor during the read-heavy workload, while the \bplustree{} is completely opposite.

Making the buffer tree adaptive introduces several challenges. First, the original buffer-tree buffers the writes and the range searches~\cite{arge2003buffer}. This degrades the latency of individual range searches. As we are dealing with oscillating workloads, batching range search queries during the read-heavy phase may not be favored from latency perspective. Second, the leaf level of the buffer tree is the same as the \bplustree{}, where data is stored in leaf pages. However, merging data from the buffers of the tree with leaf pages is expensive and can be blocking. Third, the buffer is composed of several blocks of fixed size. When emptying a buffer, this requires expensive merge-sort and disk I/Os. Lastly, straightforward adaptation of the buffer-tree overlooks the fact that most real-world data is skewed, meaning that the datasets have hotspots.


\section{Overview of the \sys{}}\label{section:overview}

To smoothly transform between the \bplustree{} and the \lsmt{}, we use an intermediate structure. Idreos et al.~\cite{idreos2019design}  propose a  continuum among indexes that indexes can be generalized under the same set of parameters. The transition between the B-tree and the \lsmt{} is bridged by the B$\epsilon$-tree~\cite{bender2015introduction} and bLSM~\cite{sears2012blsm} (\cite{idreos2019design}'s Figure 5). 
\sys{} advances this idea by adopting a hotspot-aware, tree-like architecture. Unlike traditional buffer tree~\cite{arge2003buffer} or B$\epsilon$-tree~\cite{bender2015introduction}, \sys{} uniquely integrates fine-grained adaptability, ensuring efficient transitions while maintaining high performance across varying workloads.

\begin{figure*}[ht]
    \centering
    \begin{subfigure}{0.52\linewidth}
    \centering
        \includegraphics[width=\linewidth]{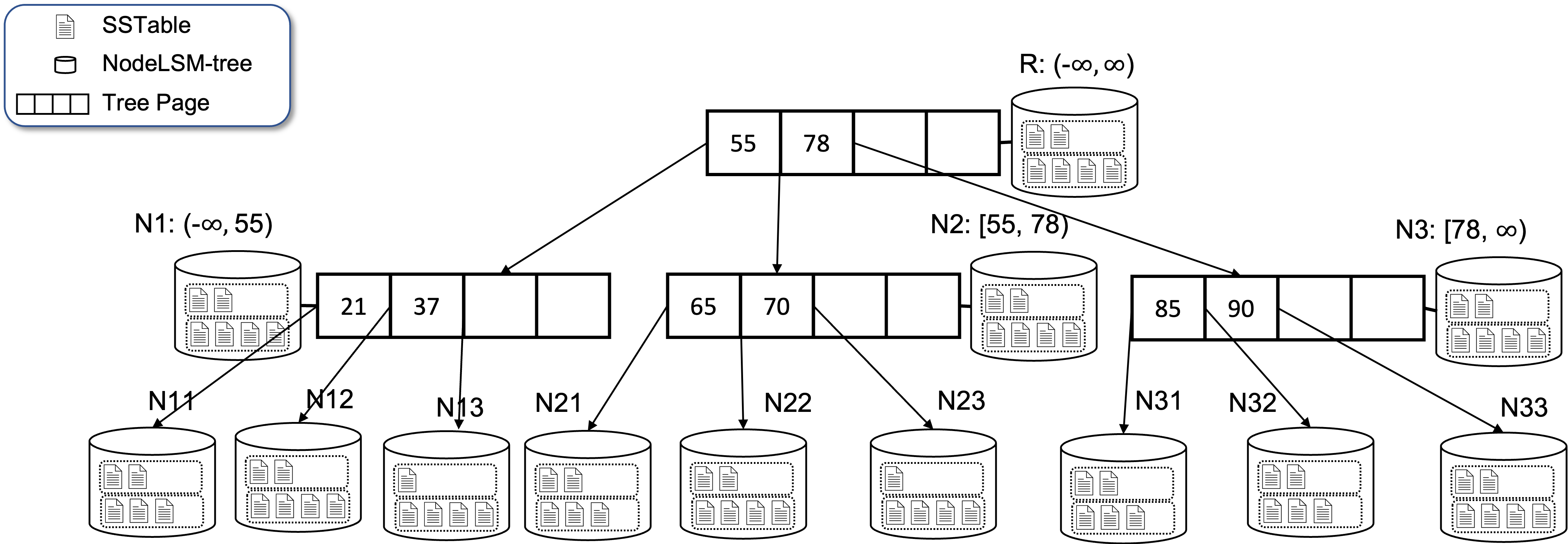}
        \caption{The basic structure of \sys{}.}
    \label{fig:aha-struct}
    \end{subfigure}
    \hfill
    \begin{subfigure}{0.47\linewidth}
    \centering
        \includegraphics[width=\linewidth]{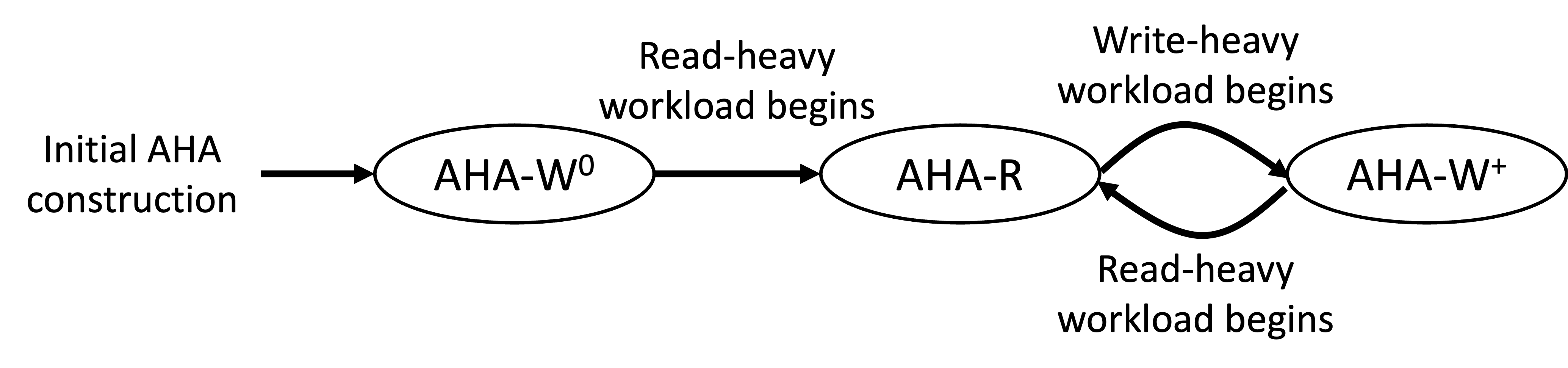}
        \caption{State transition diagram of the \sys{}.}
    \label{fig:state-machine}
    \end{subfigure}
    \caption{The basic structure of \sys{} and the life cycle of \sys{}.}
    \label{fig:aha-basic-overview}
\end{figure*}

\subsection{Basic Structure}
The \sys{} is a blend of the \bplustree{} and the \lsmt{} with a 
new procedure to maintain the tree structure. The problem in adapting between \lsmt{} and \bplustree{} is that the two structures are completely different. \sys{} counters this by introducing an intermediate structure that resembles the buffer tree~\cite{arge2003buffer} where data are stored in the \lsmtcomp{} (called buffer in buffer tree). The other component of the \sys{} is a tree structure  to facilitate search. In Figure~\ref{fig:aha-struct}, the \sys{} can be viewed in two ways. From the \lsmt{}'s perspective, all the cylinder parts (each is an \lsmt{} as well) form a big \lsmt{}. From the \bplustree{}'s perspective, all the tree pages 
form a \bplustree{}.

\subsection{Multiple States}
In order to adapt to write- and read-intensive workloads, the \sys{} transitions through multiple states over the course of its lifecycle. These states include the write-optimized \sysw{} and \syswp{}, as well as the read-optimized  \sysr{}. The state transition diagram is given in Figure~\ref{fig:state-machine}. 

Constructing the \sys{} from scratch begins in the initial write-optimized state \sysw{} where data are ingested  similar to an \lsmt{}. \sysw{} offers the fastest insertion rate among all the states as constructing an index is a long-running write-only workload.
After construction, the \sys{} is ready to handle other types of workloads, e.g., range queries. \sysw{} starts to adapt itself to \sysr{} when the workload becomes read-intensive. Since the queries focus on the hotspot, these range queries search \sysr{} as if it is a \bplustree{}.
\sysr{} can adapt to the write-optimized state in response to a write-intensive workload, transitioning to the \syswp{} state that is slightly different from \sysw{}. \syswp{} can again adapt to \sysr{} when the workload becomes read-intensive. We present details of each state in Sections~\ref{section:w0}-\ref{section:wp}, and discuss the transition process in Section~\ref{section:transition}.

\section{The Initial State: \sysw{}}\label{section:w0}

The general structure of \sysw{} is given in Figure~\ref{fig:aha-struct}. In this section, we break \sysw{} into different components, and elaborate on each of them in  Section~\ref{subsection:w0struct}.

\subsection{The Structure}\label{subsection:w0struct}

\sysw{} consists of two components: an \lsmtcomp{} and a tree structure. The \lsmtcomp{} has both an in-memory and a disk-based parts, similar to other \lsmt{} systems, e.g.,~\cite{leveldb,rocksdb}. The in-memory part includes a mutable \memtab{} and an immutable one. Writes are batched together in \memtab{} and are written to disk storage as a sequential log (termed an \sst{}). Each \sst{} contains a sorted sequence of keys, and serves as the basic unit that contains the data. \sst{}s form the disk-based part.

The tree structure is a disk-based \bplustree{} with minor modifications. A fixed-size buffer pool is used to load tree pages from disk or evict them from memory. Since data is stored in \sst{}, the leaf pages of \sysw{} remain empty as no children nodes need to be pointed to.

\begin{figure}[h]
    \centering
    \includegraphics[width=0.9\linewidth]{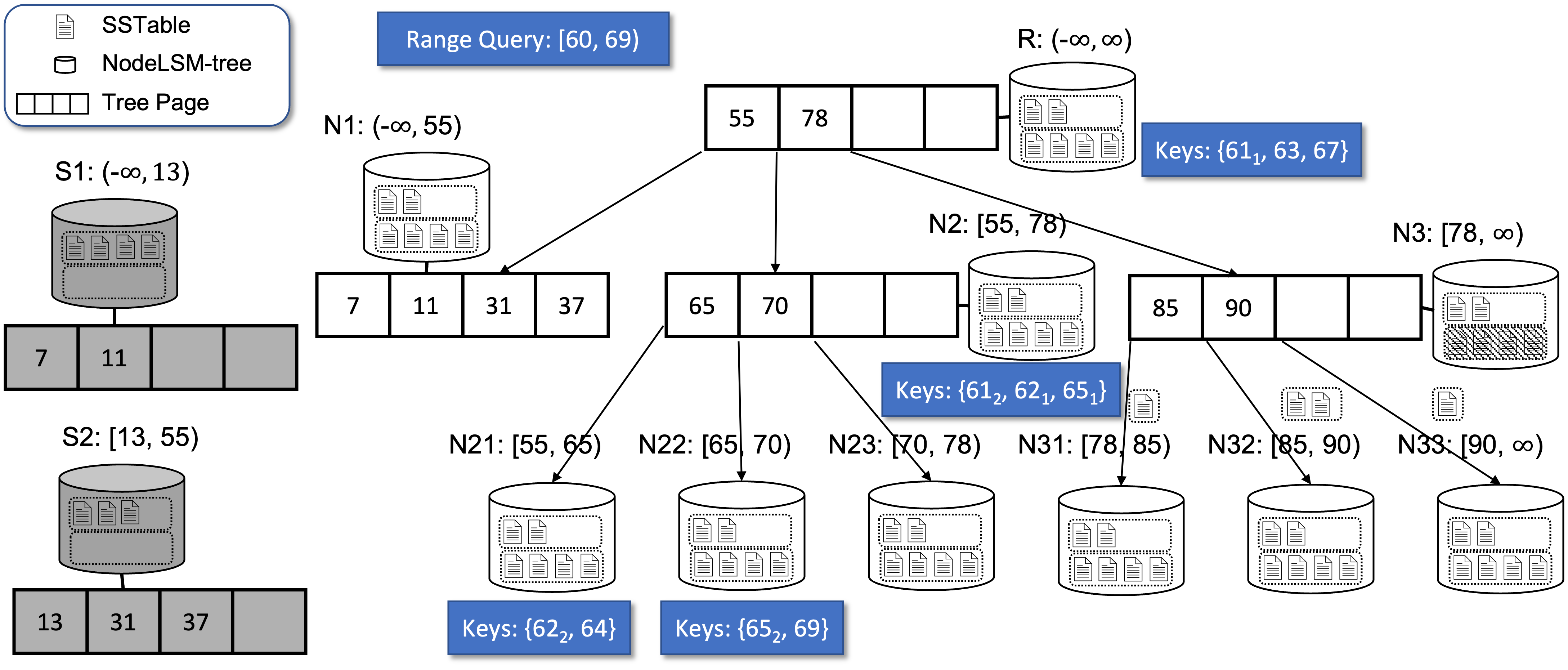}
    \caption{The structure of \sysw{} and the supported operations. The figure legends are on the top left. \sysw{} is a three-level tree structure covering from $-\infty$ to $\infty$. The nodes S1 and S2 are the results of splitting N1 when key 13 is added. A range query example is shown in the shaded text box. The Node-Emptying process of N3 sends four files to children nodes and removes them from N3's \ndl{}.}
    \label{fig:aha-w0}
\end{figure}

To integrate the \lsmtcomp{} with the tree structure, the disk-based part of \lsmtcomp{} is divided into mini-components that we term \ndl{}s. Each \ndl{} is associated with a tree page in the tree structure. Each \ndl{} is itself an \lsmt{} without \memtab{}s, representing a subset of the \lsmtcomp{}. We refer to this combination of a tree page and its corresponding \ndl{} as a \textit{Node}. The range covered by the tree page and the associated \ndl{} is consistent, forming the range covered by the Node. For example, in Figure~\ref{fig:aha-w0}, the range of Node N3 encompasses the union of the ranges of its children Nodes N31, N32 and N33. And the ranges of Nodes N31, N32 and N33 do not overlap.

The \sys{} maintains these \noindent{\bf Data Freshness Invariants}:
\begin{enumerate}
  \item Given the same key, the value in memory is fresher than the value in any of the \ndl{}s.
  \item Given the same key, the closer an \ndl{} to the root, the fresher the value.
\end{enumerate}
The invariants are enforced in all states at all times for correctness of execution.

\sysw{} without the tree structure can be seen as an \lsmt{}, but can also be viewed as a buffer tree with a much larger ``buffer" (\ndl{}, in our case). A similar, yet distinct, approach is in PebblesDB~\cite{raju2017pebblesdb}, where an \lsmt{} is blended with a skiplist to reduce  write amplification. In PebblesDB, each level is partitioned into disjoint units, separated by a set of guard keys~\cite{raju2017pebblesdb}. Similarly, RocksDB~\cite{rocksdb} employs a {\em fractional cascading} optimization~\cite{mehlhorn1990dynamic, chazelle1986fractional} that minimizes unnecessary binary searches.


\subsection{Operations}

We focus on range search queries as a representative read-heavy analytical query. In contrast to a point query, the range query selectivity can be tuned to reflect the accessed portions of data.
This way, we can simulate the analytics read-heavy workload phase using simple range searches. 

\subsubsection{Range Queries}

\begin{figure}[h]
    \centering
    \begin{subfigure}{0.49\linewidth}
    \centering
        \includegraphics[width=\linewidth]{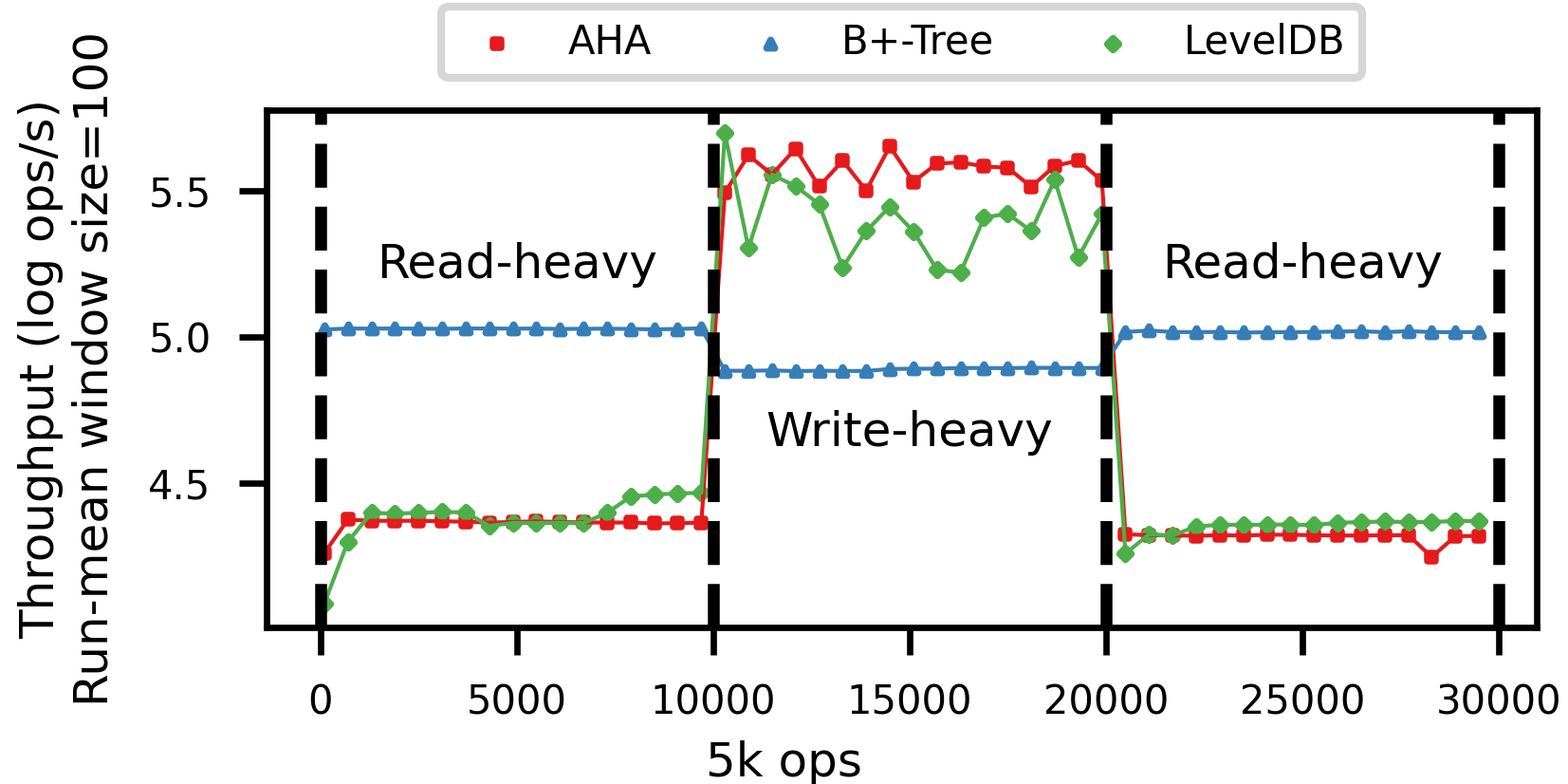}
        \caption{Throughput for uniform data.}
        \label{fig:aha-wo-perf-unif}
    \end{subfigure}
    \hfill
    \begin{subfigure}{0.49\linewidth}
    \centering
        \includegraphics[width=\linewidth]{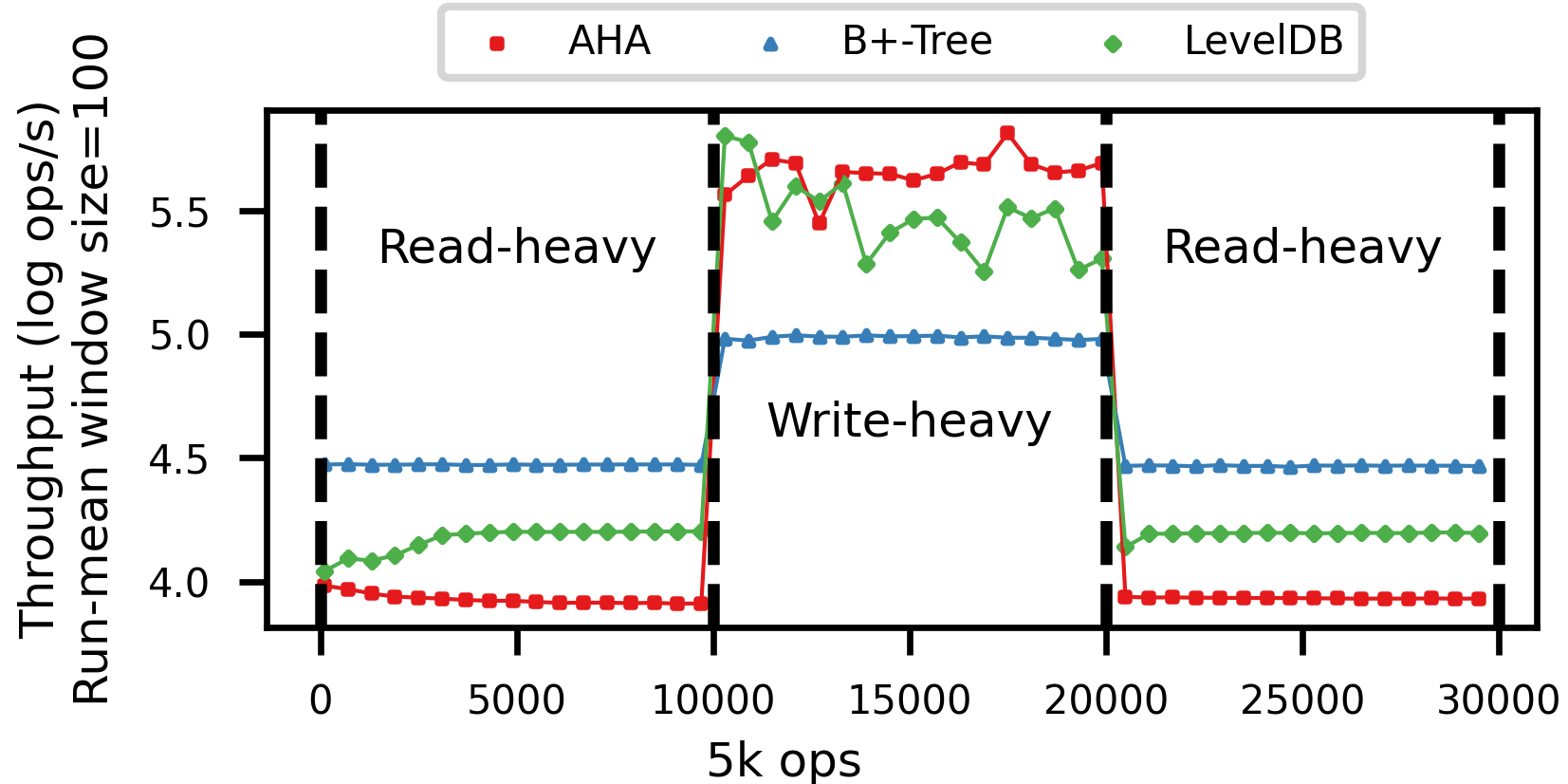}
        \caption{Throughput for Zipfian data}
        \label{fig:aha-wo-perf-zipf}
    \end{subfigure}
    \caption{Performance of \sysw{} and baselines for oscillating write- and read-heavy workloads of uniform (left) and Zipfian (right) data. \sysw{} and \lsmt{} have similar throughput for uniform data under all operations, but \sysw{} shows lower throughput for Zipfian data during read-heavy-phases.}
    \label{fig:aha-wo-perf}
\end{figure}

Based on the freshness invariant, search begins from the \memtab{}s to the \ndl{}s. If a key is duplicate, we follow the invariant and abandon the obsolete value in the result. To improve latency of the analytics operations, range  queries are not batched as is the case in the buffer tree~\cite{arge2003buffer}. For example, in Figure~\ref{fig:aha-w0}, the range query is from 60 to 69. We First, we search both mutable and immutable \memtab{}s. Then, we search  the \ndl{} of the root node as it covers the entire key space. The keys that belong to the query are \{$61_1, 63, 67$\}. Next, we navigate the search to the corresponding children nodes that overlap  the range query, i.e., Node N2, and find keys \{$61_2$, $62_1$, $65_1$\}. Finally, we reach Leaf Nodes N21 and N22 and the search stops. Keys \{$62_2, 64$\} and \{$65_2, 69$\} are reported. Obsolete values are removed based on the invariant, and the resulting values are \{$61_1, 62_1, 63, 64, 66_1, 67, 69$\}.

Figure~\ref{fig:aha-wo-perf} illustrates the throughput comparison between \sysw{} and the baselines under oscillating read-heavy (100\% read) and write-heavy (100\% write) workloads. More detail about this experiment is in Section~\ref{section:experiment}. For uniform data, \sysw{} performs similar to the \lsmt{} in all three phases, read-heavy, write-heavy and read-heavy. With a Zipfian dataset, \sysw{} exhibits  inferior read performance than the \lsmt{}.
This is due to the highly skewed nature of the Zipfian dataset, which may result in duplicate keys. Since each \ndl{} is a smaller-sized \lsmt{}, duplicate keys can persist for a longer duration from root \ndl{} to leaf \ndl{}, leading to increased overhead during queries. As \sysw{} is a write-optimized state and no adaptation is enabled, \sysw{} performs well only  in the write-heavy phase. With adaptation enabled, \sysw{} can transform to \sysr{} upon a  workload change and show better performance.

\subsubsection{Writes}
\sysw{} performs writes as if it were an \lsmt{}, treating both updates and inserts similarly by batching them in \memtab{}. All writes are first batched in \memtab{} before being written to \sst{} on disk. Each \sst{} is  added to the \ndl{}. This process occurs within the \lsmtcomp{} of \sysw{} and is the same as in other \lsmt{} implementations. Next, we  describe how the invariant is enforced in the tree structure.

Maintaining the invariant requires data to flow from memory to disk and from the root node of the tree to its leaf nodes. This data flow aligns with the \bplustree{} structure, so maintaining a valid tree structure ensures the preservation of the invariant. Each \ndl{} is an \lsmt{} that serves as a subset of the \lsmtcomp{}. Data from \sst{} travel within \ndl{}s from the top level $L_1$ to $L_{Max}$ through \lsmt{} level compaction. An overflowing \ndl{} refers to one that has data attempting to be inserted into $L_{Max+1}$. In this case,  excess data is distributed to the \ndl{}s of the child nodes. We refer to this process as \textit{node-emptying}, as it empties the \ndl{} of the node. The final destination for data is the leaf \ndl{}. Once  data in a leaf \ndl{} reaches $L_{Max+1}$, a \textit{leaf node split} is triggered. A non-leaf node splits when the tree page overflows. A node split splits both the  tree page and the \ndl{}.

\noindent
\textbf{The Node-Emptying Process.}
This process is triggered by an overflowing \ndl{} of a non-leaf node. In Figure~\ref{fig:aha-w0}, Node N3 overflows, and the overflowing \sst{}s are indicated by a texture. These \sst{}s sort merge by consulting the tree page of Node N3 (details in Section~\ref{subsection:write_opt}). The new \sst{}s are distributed into the child nodes according to the range. 
Next, we demonstrate that the freshness invariant is maintained. Within an \ndl{} that behaves as an \lsmt{}, data flows from the top level $L_1$ to the bottom level $L_{Max}$, which does not violate the invariant. Between a parent \ndl{} and a child \ndl{}, data flows only from parent to child, and not in the reverse direction. Thus, the freshness invariant is preserved during the node-emptying process.

\noindent
\textbf{Split.}
Node splits occur in both leaf and non-leaf nodes. A leaf node splits only when its \ndl{} overflows. To split an \ndl{}, the entire \ndl{} is read and is merged into a sorted sequence of \sst{}s. These new \sst{}s are then distributed to the new leaf nodes, along with a new routing key added to the parent node.

For a non-leaf node, a split occurs when the associated tree page overflows (an \ndl{} overflow triggers the node-emptying process). Suppose a new routing key, say 13, is added to N1. First, the tree page is split in half, resulting in two new pages of Nodes S1 and S2, as  in Figure~\ref{fig:aha-w0}. Next, the \ndl{} is merged into a sorted sequence of \sst{}s. These new \sst{}s are distributed according to the new ranges  in Figure~\ref{fig:aha-w0}, ensuring that the resulting nodes do not overlap in key ranges. As in the case of leaf nodes, a new routing key is added to the parent node.

\section{The Read-Optimized State: \sysr{}}\label{section:r}

\sysr{} can be viewed as a blend structure of the \bplustree{} and the \lsmt{}. We have observed that reads mainly focus on the hotspot. Thus, we proceed to transform \sysw{} so that reading hotspot data is equivalent to reading the \bplustree{}. 

\begin{figure}[h]
    \centering
    \includegraphics[width=0.9\linewidth]{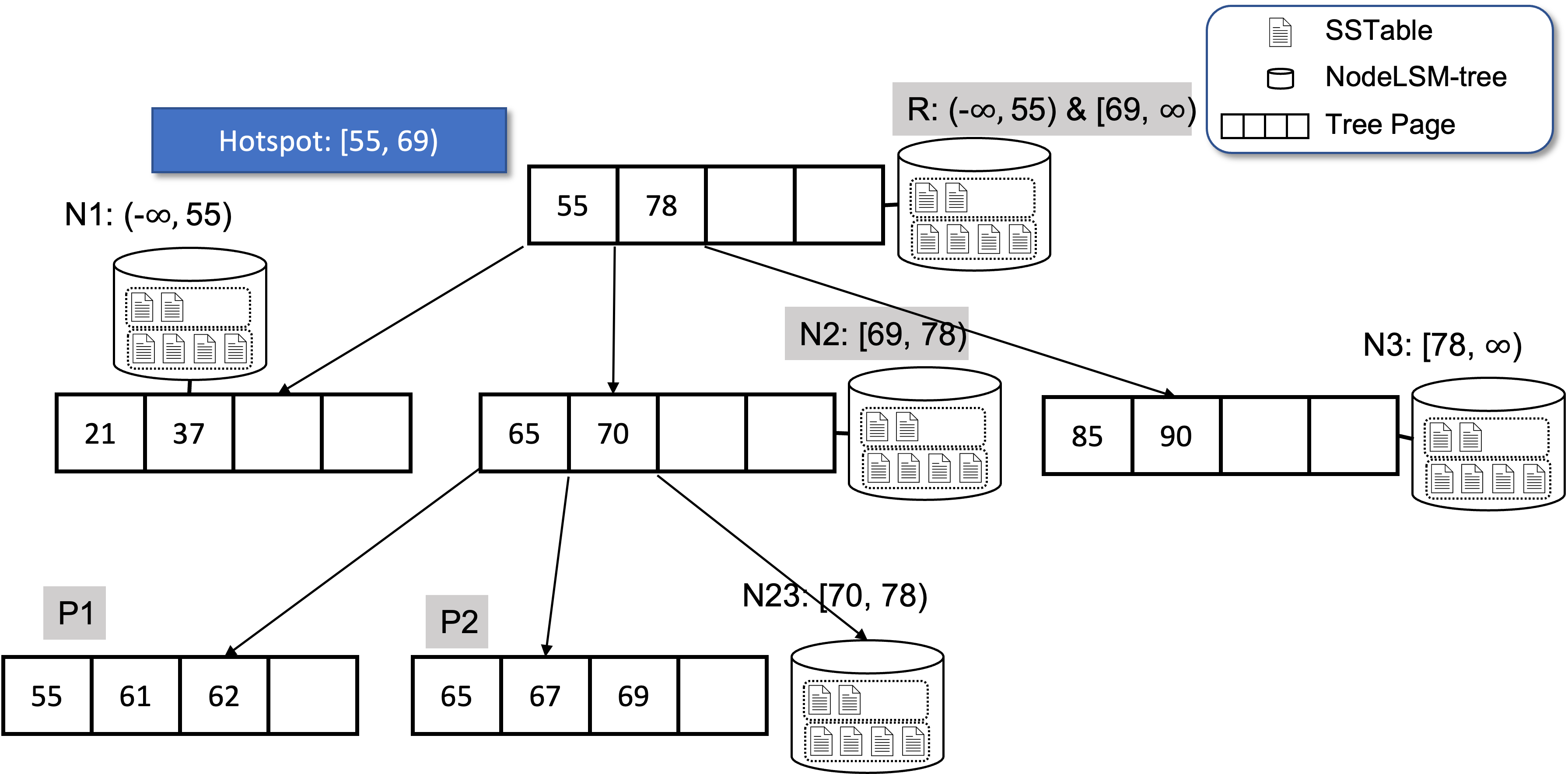}
    \caption{\sysr{} structure. The hotspot range is  55-69. Both root and N2 \ndl{}s do not contain the hotspot data.}
    \label{fig:aha-r}
\end{figure}

\subsection{The Structure}
The overall structure of \sysr{} is similar to \sysw{}. The two components: \lsmtcomp{} and the tree structure still co-exist. In \sysr{}, to facilitate reads, the leaf level within the hotspot contains no \ndl{}s. Data on the leaf level is only stored in the tree structure, i.e., in  tree pages. This reduces the search time within \ndl{} as tree pages store data sequentially. However, reads still have to probe the \ndl{}s in non-leaf levels. To further reduce the overhead of reads, we make those \ndl{}s \textit{hotspot-free}, i.e., they do not contain hotspot data. The freshness invariant is  valid in \sysr{} as all  hotspot data is stored in the leaf level only.

In Figure~\ref{fig:aha-r}, the hotspot range is [$55, 69$]. 
The root \ndl{} only covers ranges ($-\infty$, 55) $\cup$ [69, $\infty$). For non-leaf nodes N2, its \ndl{} covers range from [69, 78). All the data  inside the hotspot is located in the leaf level in tree pages. From the perspective of the hotspot range, \sysr{} behaves as a \bplustree{}. 

\subsection{Operations}

\subsubsection{Range Queries}
A range query is divided into two parts: a hotspot subquery and a cold spot subquery. Processing a cold spot subquery is the same as range querying over \sysw{}. The search starts from the \memtab{} and traverses the tree structure while reading the necessary \ndl{}s. Searching for a hotspot subquery is the same as searching the \bplustree{}. 
The subquery traverses the tree and finds all  leaf pages that overlap  the subquery. Hotspot subquery results are returned from a sequential read of these pages. In Figure~\ref{fig:aha-r}, a range query [65, 73) consists of a hotspot subquery [65, 69) and a cold spot subquery [69, 73). The final result of the range query is the combination of the above two subqueries that query pages P1 and P2, as well as the \ndl{}s of R, N2 and N22.

\begin{figure}[h]
    \centering
    \begin{subfigure}{0.49\linewidth}
    \centering
        \includegraphics[width=\linewidth]{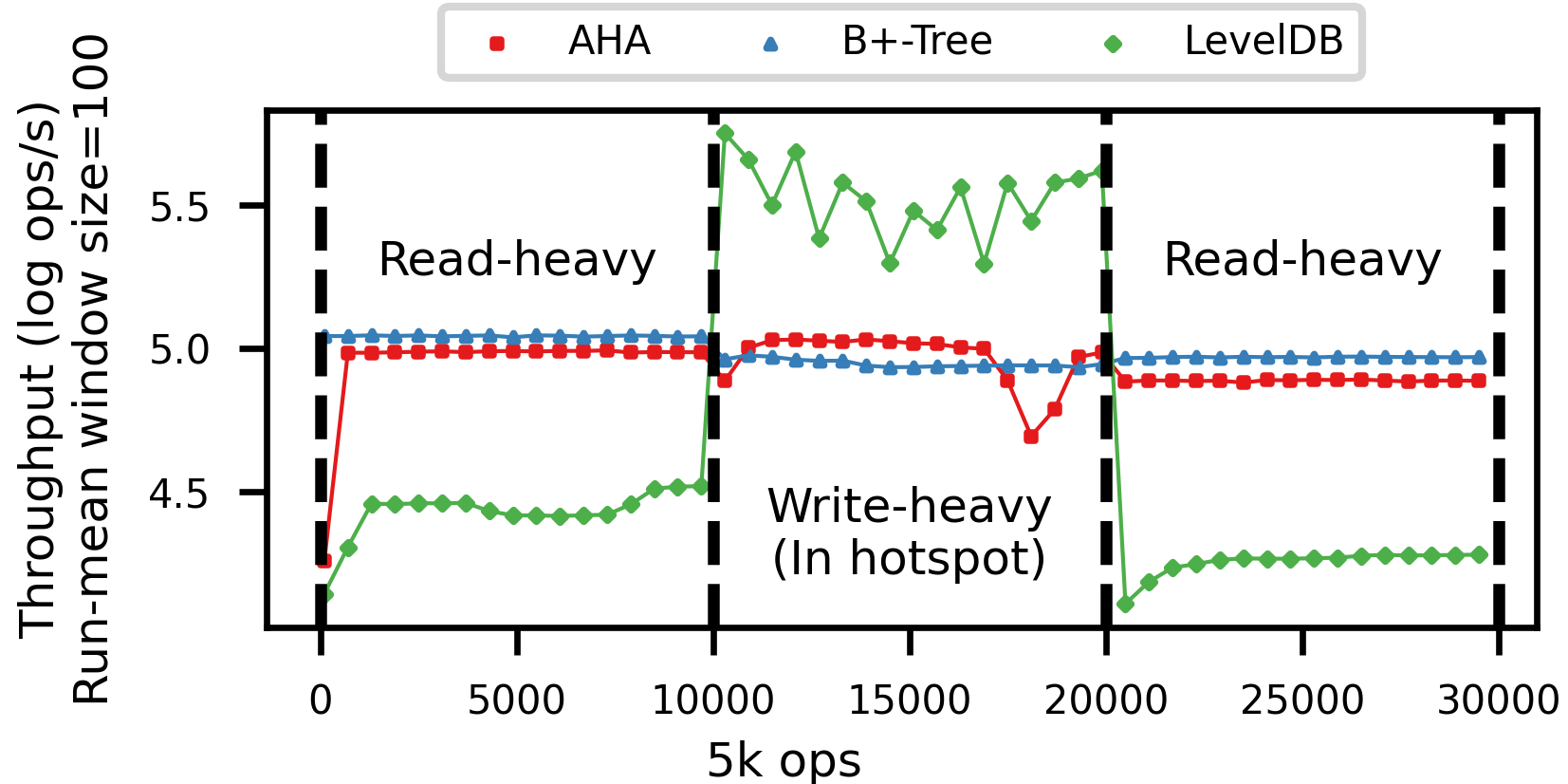}
        \caption{Update operations are within the hotspot range.}
        \label{fig:tree_insert_hot_w}
    \end{subfigure}
    \hfill
    \begin{subfigure}{0.49\linewidth}
    \centering
        \includegraphics[width=\linewidth]{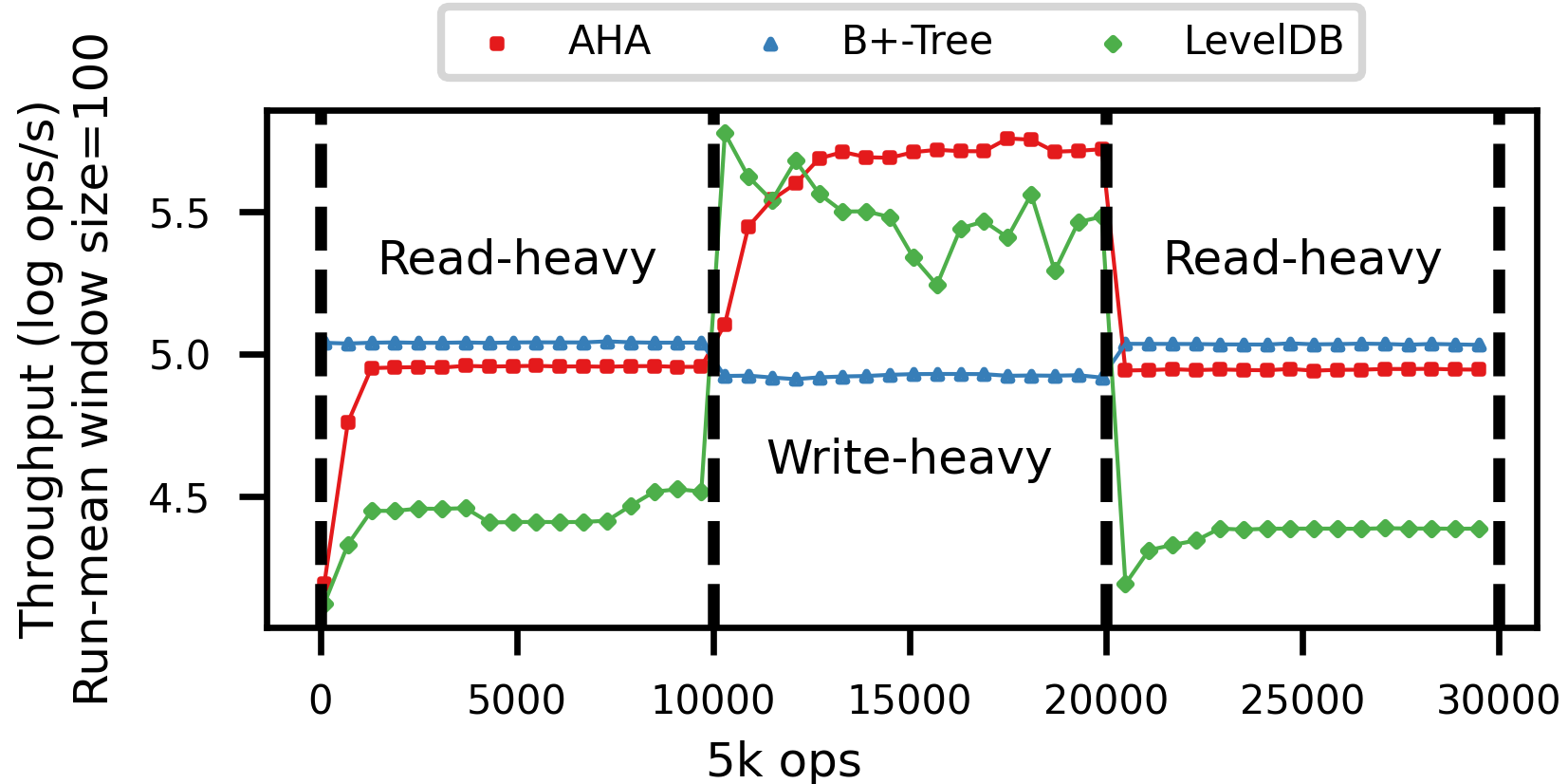}
        \caption{Update operations are spread over entire key range.}
        \label{fig:tree_insert_all_w}
    \end{subfigure}
    \caption{The performance of \sysr{} under oscillating read-heavy and write-heavy workloads.}
    \label{fig:tree-insert}
\end{figure}

\subsubsection{Writes}
Writing in \sysr{} can also be divided into two parts: a hotspot write and a cold spot write, as in the case of range queries. The reason is that hotspot data need to be kept in leaf pages only, else the range query may return incorrect results. A cold spot write is similar to a write operation of \sysw{}. Writes are batched in memory and are written to disk as \sst{}s. The tree structure and the \ndl{}s are maintained in the same way as \sysw{}. A hotspot write cannot be batched in memory as \sysr{} guarantees all hotspot data are in the leaf pages. Thus, hotspot data writes directly into leaf pages. The search of the target leaf node starts at the root node, following the \bplustree{} lock-coupling technique. Insertion is performed once the target leaf node is found. If the leaf node splits due to overflowing, a new routing key is added to the parent node. An overflowing non-leaf node is treated the same way as in \sysw{}.

In Figure~\ref{fig:tree-insert}, we give the performance of \sysr{} under oscillating read-heavy (100\% read) and write-heavy (100\% write) workloads. More detail about this experiment is in Section~\ref{section:experiment}. The difference of the two figures lies in the update operations that happen in the 50-100 Million operations range (write-heavy phase). Figure~\ref{fig:tree_insert_hot_w} gives the result of updates in the hotspot. \sysr{} has almost identical throughput to the \bplustree{} for that operation range. For Figure~\ref{fig:tree_insert_all_w}, the throughput of \sysr{} write increases gradually and remain high.
The \sys{} starts as \sysw{} at Operation 0 and transitions to \sysr{} during the first read-heavy phase (when the throughput becomes comparable to \bplustree{}). For this experiment, the \sys{} remains as \sysr{} for the remaining operations and does not transition to any other state.

\section{The Write-Optimized State: \syswp{}}\label{section:wp}
Once \sys{} is in the \sysr{} state, under the write-heavy workload, writes still need to be performed as if an \lsmt{}, resulting in State \syswp{}. \syswp{} is different from \sysw{} as \syswp{} transitions from State \sysr{}.

\begin{figure}[h]
    \centering
    \includegraphics[width=\linewidth]{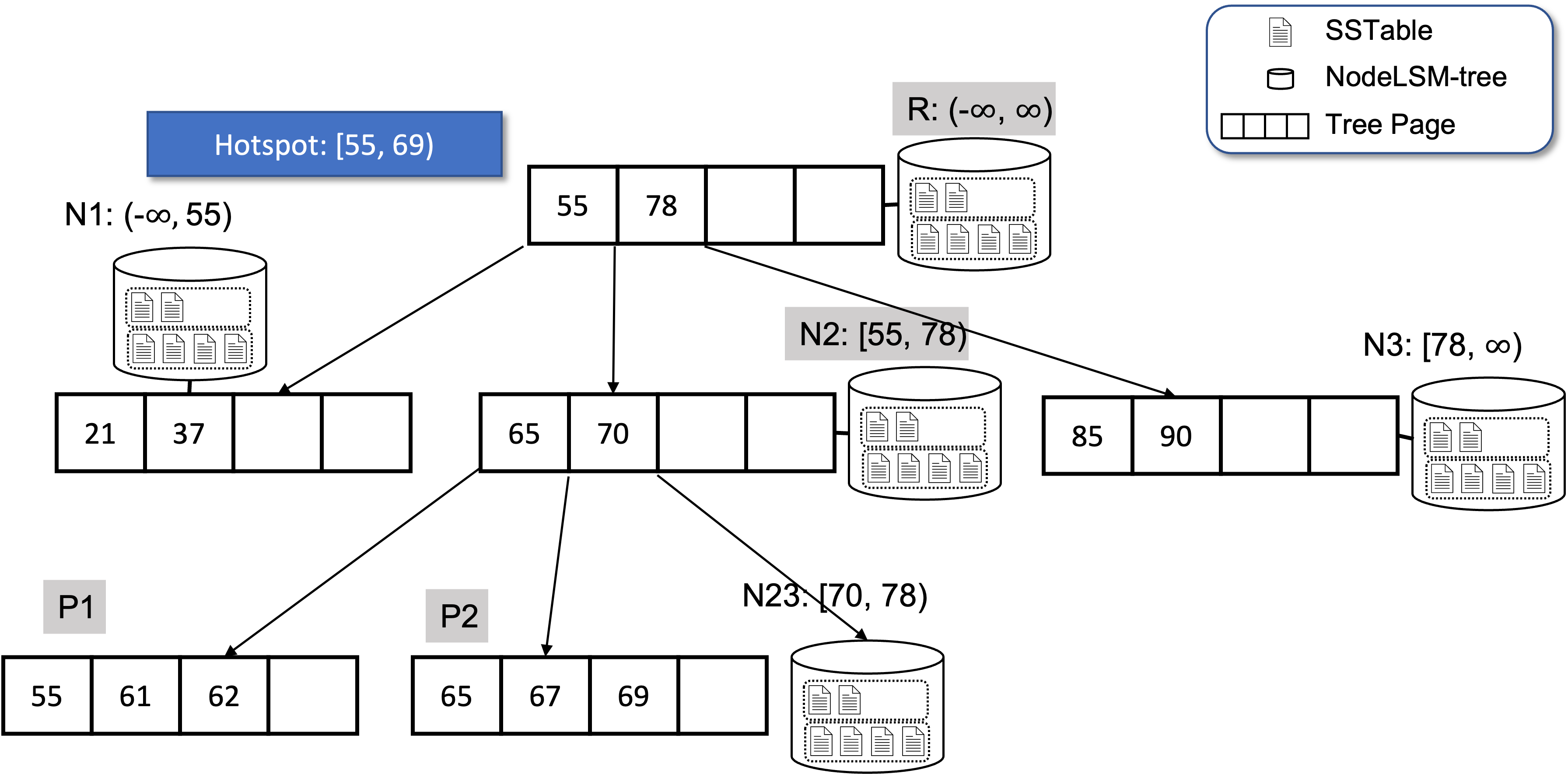}
    \caption{\syswp{} structure. The hotspot range is 55-69. Root and N2 \ndl{}s are no longer hotspot-data-free.}
    \label{fig:aha-wp}
\end{figure}

\subsection{The Structure}
The structure of \syswp{} is almost identical to \sysr{} except in the placement of  hotspot data. Hotspot data can exist in \ndl{}s of non-leaf nodes as well as in the leaf nodes.
This maximizes the write throughput of \sys{} as all data (hot or cold) are inserted in batches.
The freshness invariant is still valid as fresher hotspot data is in the closer-to-root \ndl{}s.

\subsection{Operations}

\subsubsection{Range Queries}
A range query queries both \ndl{}s as well as the leaf pages. In Figure~\ref{fig:aha-wp}, Range Query [60, 66) is within the hotspot range [55, 69). As nodes R and N2 are no longer hotspot-data-free, the process is similar to searching in \sysw{} where multiple \ndl{}s R, N2 and leaf pages P1, P2 are searched because of overlapping ranges.

\subsubsection{Writes}
In order to accommodate the write-heavy workload, \syswp{} behaves as an \lsmt{}. Writes are batched before being written to disk as \sst{}s, including both hotspot and cold spot data. However, this comes at a price of breaking the hotspot guarantee of \sysr{} where hotspot data only resides in leaf pages. In Figure~\ref{fig:aha-wp}, all the newly added data is accumulated in \ndl{}s before being flushed to either leaf \ndl{}s (cold spot data) or being merged with leaf pages (hotspot data).

\section{Optimizations}\label{section:implementation}

We observe bottlenecks in the 
current design of the \sys{}
with respect to the write throughput and  the node-emptying process. In this section, we analyze these bottlenecks,
and devise optimizations to address each of them.


\begin{figure}[h]
    \centering
    \begin{subfigure}{\linewidth}
    \centering
        \includegraphics[width=0.85\linewidth]{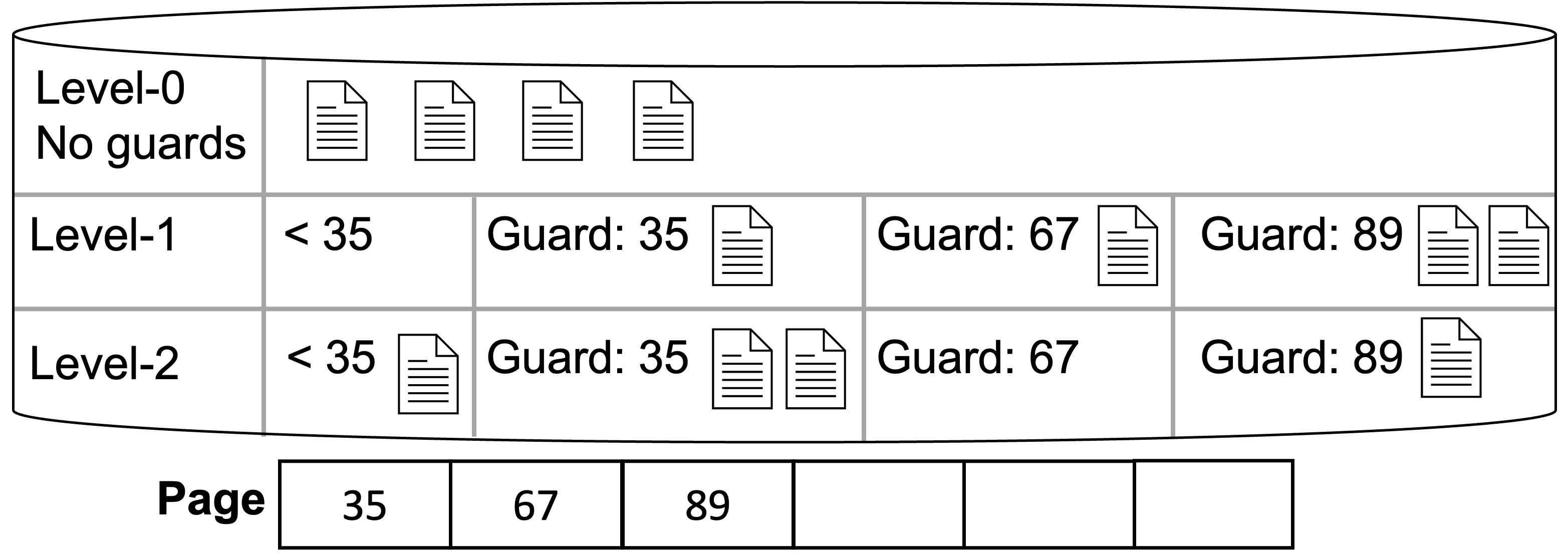}
        \caption{Guarded compaction}
        \label{fig:guarded}
    \end{subfigure}
    \vfill
    \begin{subfigure}{\linewidth}
    \centering
        \includegraphics[width=0.85\linewidth]{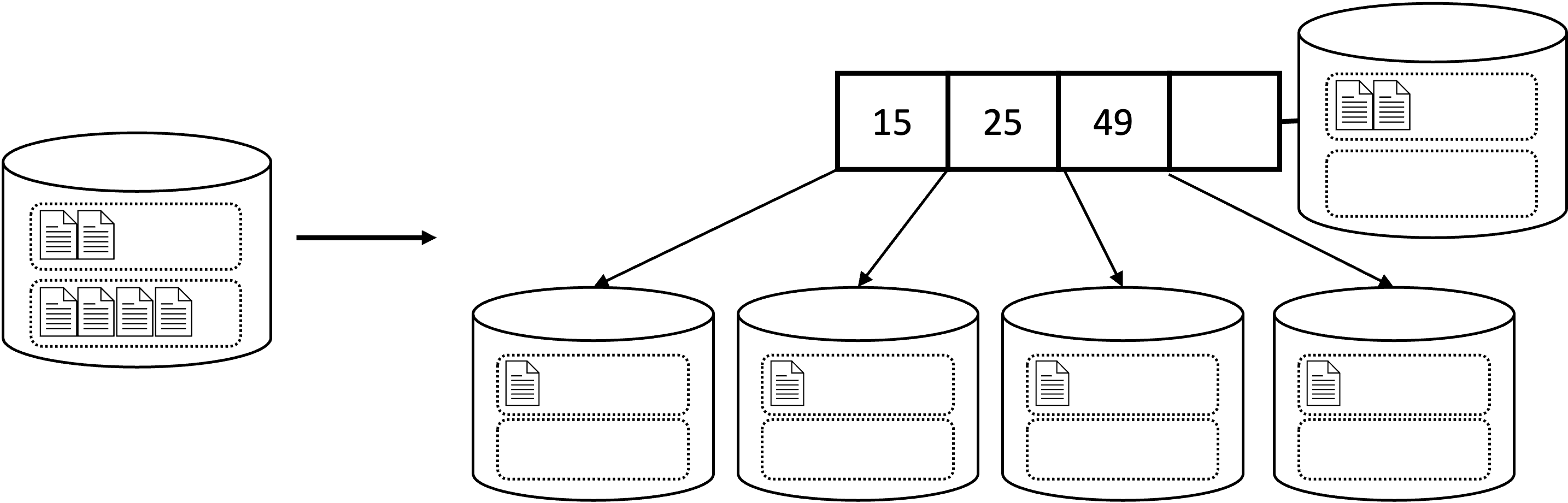}
        \caption{\sst{} distribution}
        \label{fig:file-as-node}
    \end{subfigure}
    \caption{Two optimizations to improve \lsmt{} compaction and bulk-loading time.}
    \label{fig:comp-opt}
\end{figure}

\subsection{Write Throughput}\label{subsection:write_opt}

\noindent
\textbf{Compaction Threads.}
LevelDB~\cite{leveldb} uses one background thread for \sst{} compaction. The \sys{} adopts this design. However, it is not feasible for each \ndl{} to have a dedicated background thread. Thus, only the root \ndl{} is given a dedicated background thread \verb|BGroot|,
and we refer to the  root's \lsmt{} by \rtl{}. 
Another difference in  \rtl{} is that in contrast to the other \ndl{}s, \rtl{} accepts new 
\sst{s} that result from writing \sys{}'s \memtab{} to disk.
Delaying  internal compaction of \rtl{} causes delay in accepting new \sst{}s that in turn blocks the user writes. \verb|BGroot|  alleviates the blocking caused by \rtl{'s} internal compaction.

To maintain the tree structure and execute the node-emptying process, a second background thread named \verb|BGtree| is used. \verb|BGtree| is 
invoked
by \verb|BGroot| whenever \rtl{} is ready to send \sst{}s to child nodes. \verb|BGtree| is responsible for maintaining the tree structure, including node-emptying, \ndl{} internal compaction, splitting, etc. As an overflowing \rtl{} may still be blocked waiting for \verb|BGroot| to finish, we allow a \textit{soft} size limit on \rtl{} such that the number of levels in \rtl{} can exceed the limit temporarily. This way, the incoming writes are not blocked.

\noindent
\textbf{Reducing \lsmt{} Compactions.}
During the node-emptying process, 
\sst{}s
within each \ndl{} may not be aligned with the routing keys of the associated page. As \sst{}s are immutable the same way as in LevelDB, the \sst{}s that span across routing keys cannot be added to the child nodes as is. The \sst{}s need to be rewritten into new \sst{}s, which is expensive.
This can be avoided by consulting the routing keys during internal compaction of the \ndl{}s. 
PebblesDB~\cite{raju2017pebblesdb} introduces the notion of guards, and enforces the file range within guards.
The \sys{} adopts this design, and we term this process \textit{guarded compaction}.
When a \ndl{} starts  internal compaction, the node page is used as  input, and the resulting \sst{}s are aligned with the routing keys as  in Figure~\ref{fig:guarded}. Later, during a node-emptying process, if the node page has not been modified since, the \sst{s} in the bottom level of this \ndl{} can be dispatched to child nodes without being rewritten. Thus, in most cases, \sst{}s can be migrated and rerouted directly from the bottom level of a parent node's \ndl{} to the top level of a child node's \ndl{} by pointer shuffling without reading the \sst{}s into memory.

Splitting a leaf node can be expensive as it involves reading its \ndl{}, sorting all  data entries, and writing into two \ndl{s} with non-overlapping ranges. As reading and writing \sst{s} is  expensive, to reduce the number of \ndl{} re-compactions, we split each leaf node into more than two siblings, depending on the actual number of \sst{}s in the \ndl{}. This delays the leaf node's split time but in turn may cause the non-leaf node to be full quickly. As this non-leaf node has a relatively small \ndl{} with parts of its  \sst{}s flushing down to children, splitting a non-leaf node is less expensive. 

\noindent
\textbf{Reducing Bulk-loading Time.}
Initially, \sysw{}  only contains a \rtl{} that is an \lsmt{}. To build a tree structure when \rtl{} overflows,
i.e., data is attempting to be inserted into $L_{Max+1}$ that is beyond the pre-defined level limit,
the straightforward idea is to split \rtl{} into two new leaf nodes, and insert a separator key into a new root node. 
This is time-consuming. We leverage the fact that \sst{}s within levels that are higher than Level-0 are sorted in the \lsmt{}. Instead of merging \sst{}s, we assign each \sst{} to a new \ndl{}, creating multiple leaf nodes  as  in Figure~\ref{fig:file-as-node}. In this case, no compaction or I/O is needed  as we only read the metadata of the \sst{}, 
and create a root node with \sst{} ranges as the routing keys. Also, we apply this technique to the state transitioning process from \sysw{} to \sysr{} (Section~\ref{section:conclusion}).

\subsection{Concurrency}
\noindent
\textbf{Double Buffering.}
It is challenging to enable concurrency in the \sys{}. Optimistic lock coupling (OLC) may not be suitable. The reasons are twofold. First, \rtl{} is frequently modified in states \sysw{} and \syswp{}, which add \sst{}s to \rtl{}. Second, \rtl{} includes keys that spread all over the key space, and is involved in all read operations. If OLC were applied in the \sys{}, there is high likelihood of frequent read validation failures on  \rtl{}. 
Another way to achieve concurrency is to use a lock-free structure. As in the Bw-tree~\cite{levandoski2013bw}, the intermediate structure is valid via atomic installation of small changes. For the \sys{},  atomic installation becomes complicated as there can be changes both to the tree page as well as to the \ndl{}. Moreover, the node-emptying process requires changes on multiple nodes at the same time (one parent node and multiple children nodes). 
In our implementation, we lock this 
subtree to enable atomic changes.
However, locking the subtree is  blocking. Moreover, 
\ndl{} compaction is a long and time-consuming process. To resolve this issue, we use \textit{double buffering} in  \ndl{}. We allow at most two versions of the \ndl{} for one node at any time. The version visible to the reader is the one before any modifications can happen. The other version only exists when \verb|BGtree| is compacting or preparing for the node-emptying process. Only the \ndl{} structure is duplicated but the \sst{}s are not copied. \verb|BGtree| compacts in the background without locking. Changes installed atomically to the tree become visible to readers only after \verb|BGtree| has completed. Obsolete \sst{}s and useless nodes are removed when no readers can see them. Node split is achieved in a similar fashion that the readers continue to read the old-versioned subtree until \verb|BGtree| installs the new subtree atomically.

\section{Transitions Among The \sys{} States}\label{section:transition}

As  in Figure~\ref{fig:state-machine}, the \sys{}  transitions from the write-optimized state \sysw{} to read-optimized \sysr{}, as well as between State \sysr{} and the write-optimized state \syswp{}. Sections~\ref{subsection:w2r} and \ref{subsection:r2w} describe this process.

\subsection{Transitions To The Read-Optimized State}\label{subsection:w2r}

\begin{figure}[h]
    \centering
    \includegraphics[width=0.9\linewidth]{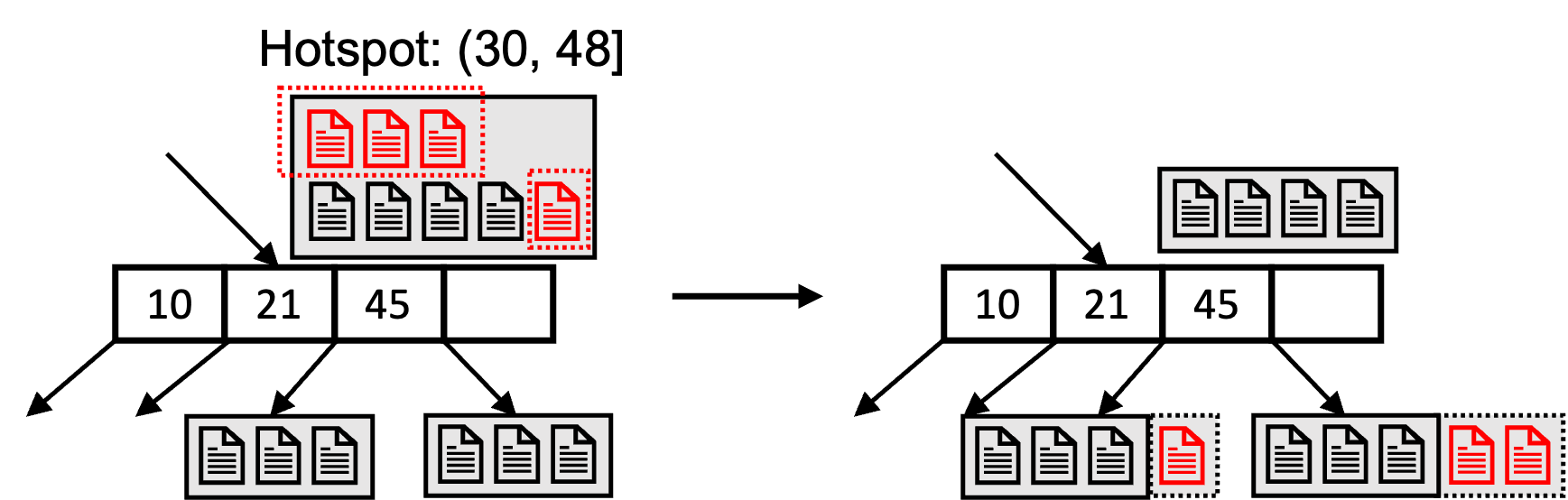}
    \caption{The Hotspot-Emptying process. Only the \sst{}s that overlap the hotspot range are compacted and are sent to the child \ndl{}s.}
    \label{fig:hotspot-empty-process}
\end{figure}

\subsubsection{\sysw{} to \sysr{}}\label{sssection:w2r}
As the workload shifts from being write-heavy to being read-heavy, \sysw{} adapts so that the resulting 
index is read-optimized. 
Automatic identification of hotspots is orthogonal to our study.  In this paper, we assume that hotspots are detected online and are known. Many research works identify hot from cold data, including~\cite{anneser2022adaptive,kakaraparthy2022vip,zhang2022sa,ge2023sali}. Once a hotspot is detected,  we describe below how the adaptation process takes place.

The adaptation process is conducted by the background thread \verb|BGtree|, the same thread that is responsible for  tree structure maintenance. For read-heavy workloads, the goal is to have all hotspot data entries be stored in the leaf pages of \sysr{}. Reading only the leaf pages avoids searching for data inside \rtl{} or \ndl{}s, which is expensive. The whole process involves a \textit{Hotspot-Emptying Process} and a \textit{Leaf \ndl{} Transformation Process}.

\noindent
\textbf{Hotspot-Emptying Process}. One brute-force method to avoid searching a \ndl{} is to send all data entries in \ndl{} to the leaf pages. However, real datasets always have hotspots and focusing solely on hotspot data is more efficient. To migrate hotspot data from \rtl{} and \ndl{}s, all the tree nodes from root to leaf having overlapping ranges with the hotspot range need to be checked. The adaptation of a node \ndl{} is query-triggered, i.e., we examine the \ndl{}s if these nodes are queried. During range search, if the tree traversal discovers an \ndl{} that overlaps with the hotspot, \verb|BGtree| is waken to compact all the data items that belong to the hotspot, and then sends those \sst{}s to the children nodes as  in Figure~\ref{fig:hotspot-empty-process}. The original \ndl{} becomes hotspot-free, i.e., this \ndl{} does not own any hotspot data. We repeat this process and the resulting \sysr{} is the one  presented in Figure~\ref{fig:aha-r} non-leaf levels. This process is almost identical to the node-emptying process except that the way to gather the \sst{}s is different.

\begin{figure}[h]
    \centering
    \includegraphics[width=0.95\linewidth]{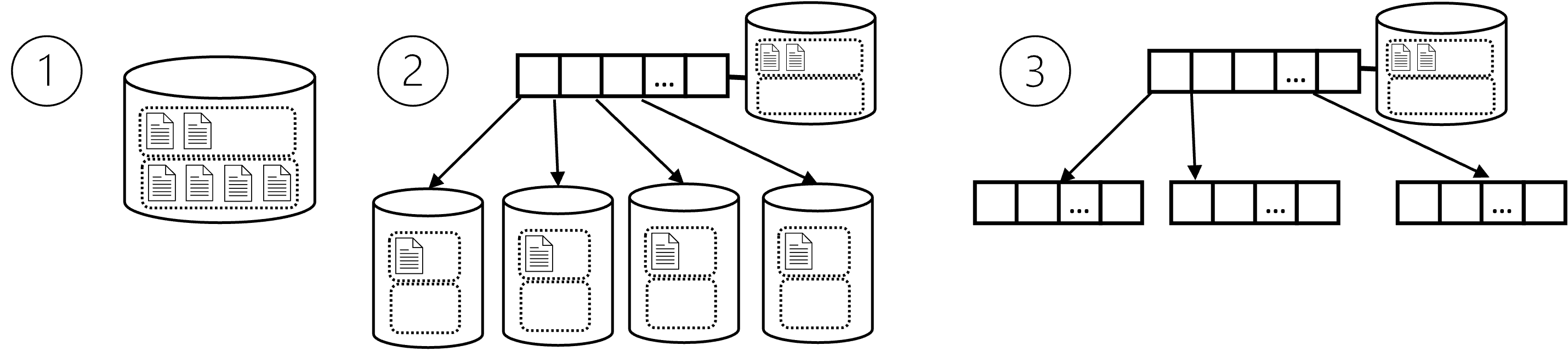}
    \caption{Leaf \ndl{} transformation process}
    \label{fig:leaf-transform}
\end{figure}

\noindent
\textbf{Leaf \ndl{} Transformation Process}. Leaf nodes with \ndl{}s are transformed into regular \bplustree{} leaf pages. 
A straightforward way is to read the entire \ndl{} into memory, sort it, write the data entries in leaf pages, and discard the original \ndl{} (we term it a  \textit{side-split}). However, as  in Section~\ref{section:implementation}, this is an expensive process. To speed it up, we apply the same technique in Section~\ref{section:implementation}, and break the process into two steps to avoid substantial modifications to the tree structure (we term it \textit{down-split}). The first step is to rewrite leaf \ndl{} into sorted single-\sst{} \ndl{}s as  in Figure~\ref{fig:leaf-transform} \circled{2}. 
In the following step, we transform this single \sst{} leaf node by reading 
it and 
writing 
its
data into new leaf pages and removing the old \sst{}.

In the first step, the bottom level of this leaf's \ndl{} is extracted and the routing keys are added to the tree page (the original leaf). The original leaf node becomes a non-leaf node with the remaining \sst{}s in its \ndl{}. This makes the tree unbalanced as it adds one more level in the middle but does not require any compaction during split. 

The benefits are twofold. First, we avoid the lengthy process to compact \ndl{}. Second, growing the node depth avoids the overhead of splitting non-leaf nodes. Since tree pages are in smaller sizes than \sst{}s, the resulting number of pages can be large. Thus, the number of inserted routing keys in the parent node is large. A tree page that is over 50\% full is more likely to overflow than a completely empty page. By introducing one more node along the path, we reduce the number of non-leaf node splits. 

\subsubsection{\syswp{} to \sysr{}}
Instead of \ndl{}, \syswp{} already has tree pages in the leaf level within hotspot. Thus, there is no need to undergo the leaf \ndl{} transformation process. When the hotspot-emptying process is triggered, the hotspot data residing in \ndl{}s is sent down until the leaf level, and gets merged with the existing leaf pages.

\subsection{Transition To The Write-Optimized State}\label{subsection:r2w}
Transitioning from \sysr{} to \syswp{} is trivial. When the workload becomes write-heavy again, data are batched in memory to speed up the writes and the existing \sys{} becomes \syswp{}.

\section{Experiments and Analysis}\label{section:experiment}
In this section, we evaluate the \sys{} against other indexes under oscillating write-heavy and read-heavy workloads, and analyze the performance.

\subsection{Experimental Setup}
We use a 152-core Intel(R) Xeon(R) Platinum 8368 CPU @ 2.40GHz with 197 GB memory running Ubuntu 22.04.2 LTS of two NUMA nodes. We pin all our experiments in one NUMA node to eliminate NUMA-related performance issues. All tested indexes are disk-based: the \bplustree{} and the \sys{} are equipped with an in-house buffer manager.

We use two data distributions in the experiments: uniform distribution and Zipfian distribution to mimic real datasets. The key for a data item is a 20-byte string, and the value is a 128-byte string. All  indexes are loaded by being inserted with 500 Million unordered key-value pairs. The hotspot range for the uniform dataset is of size 10\% of the entire key-space, and 1\% for the Zipfian dataset. The hotspots start at the leftmost value of the key-space. The range search query covers $2\times 10^{-4}$\% of the total key space for the uniform dataset and $2\times 10^{-5}$\% for the Zipfian dataset. After index construction, we simulate the oscillating read-heavy and write-heavy workloads by issuing 50 Million range search queries, followed by 50 Million updates and then 50 Million range search queries. We use 100\% read (write) operations during each oscillating phase to stress-test the indexes. The throughput is recorded starting from Operation 0 that corresponds to the first operation in the workload. The plotted throughput is presented on a logarithmic scale and averaged using a running mean window of size 100, unless otherwise indicated.

\subsection{Indexes Under Comparison}

The \sys{} adopts the \lsmt{} implementation of LevelDB~\cite{leveldb} as an initial structure and modifies on it. Thus, we compare the \sys{} with LevelDB in the experiments. However, since the implementation of the \lsmtcomp{} is pluggable for the \sys{}, other \lsmt{} can fit as well. We implement an in-house disk-based \bplustree{} for the experiments. During the experiments, user threads send requests, including write operations and range search queries, to the index. The \sys{} uses two background threads during index modification  and LevelDB uses one thread. 
For a fair comparison with the \bplustree{} and the \lsmt{}, we maintain the same number of threads running for each index at any given time (the sum of the user threads and background threads is the same
e.g., the \bplustree{} uses three user threads, LevelDB uses two user threads plus one background thread, and \sys{} uses one user thread plus two background threads). If one background thread finishes its work, one user thread can be awakened to execute the available operations.

\subsection{Effectiveness of Adaptation}
\begin{figure*}[h]
    \centering
    \begin{subfigure}{0.22\linewidth}
    \centering
        \includegraphics[width=\linewidth]{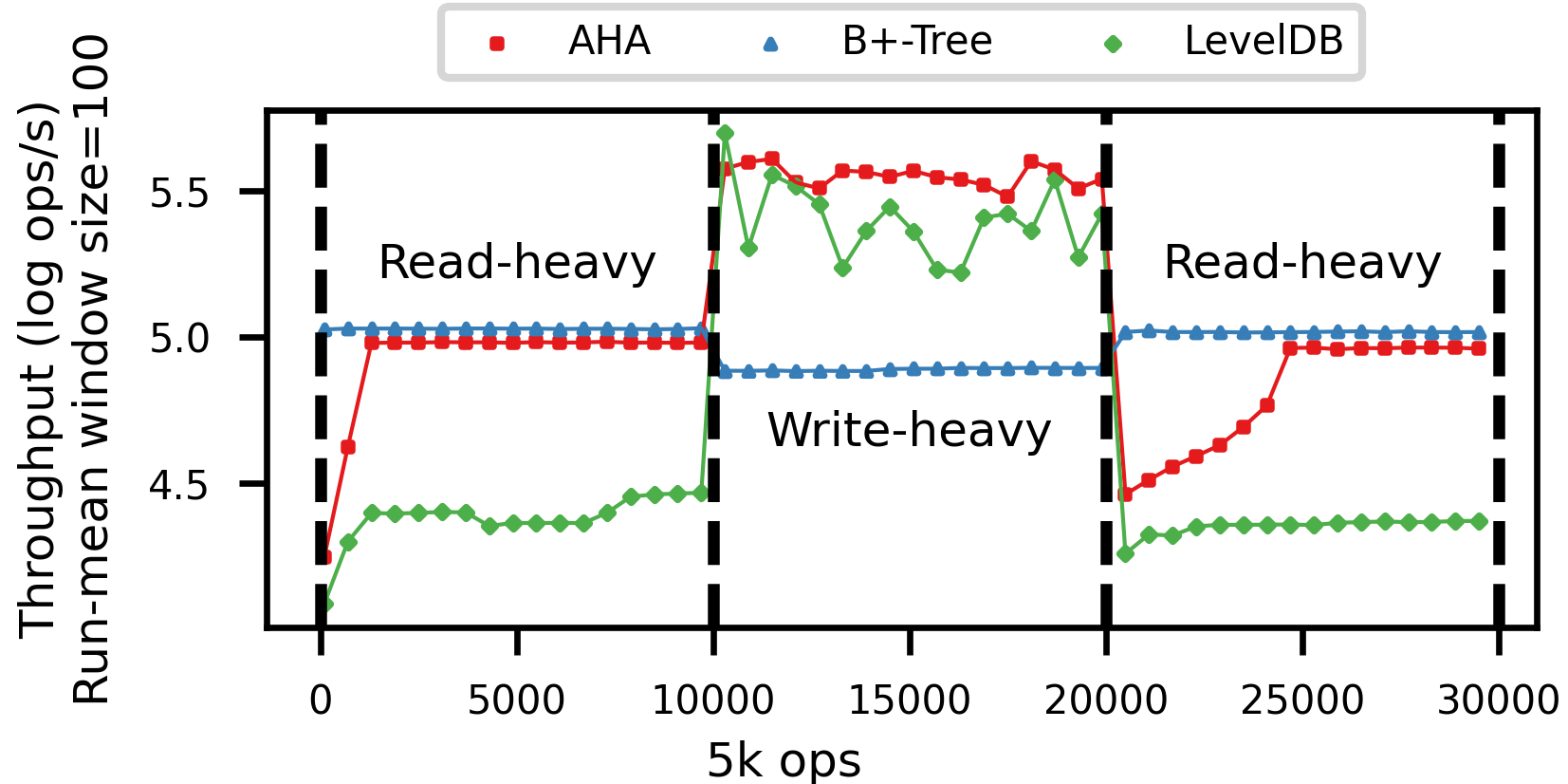}
        \caption{Uniform dataset with oscillating workloads.}
        \label{fig:unif_baseline}
    \end{subfigure}
    \hfill
    \begin{subfigure}{0.22\linewidth}
    \centering
        \includegraphics[width=\linewidth]{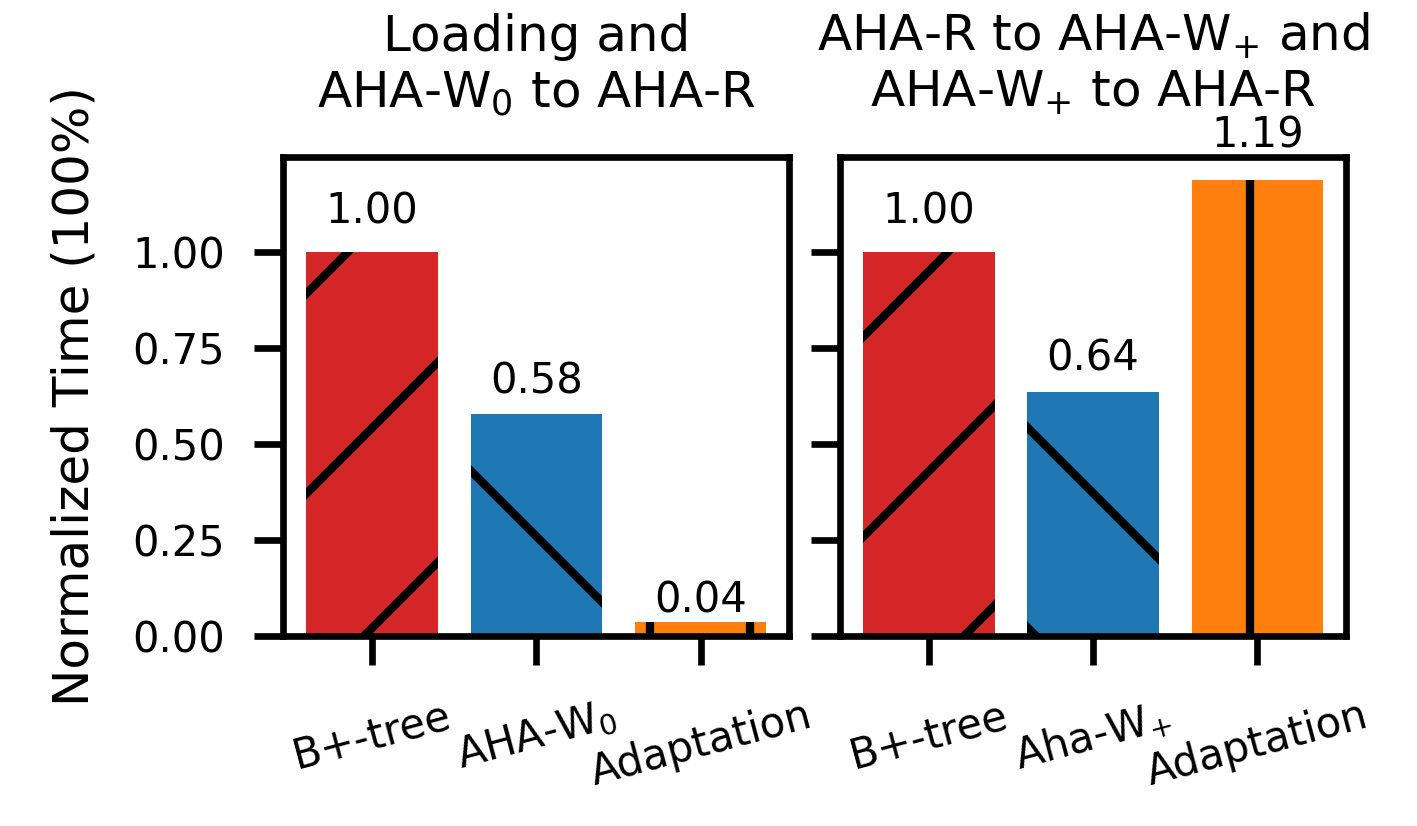}
        \caption{The elapsed adaptation time (Uniform dataset).}
        \label{fig:unif_construct}
    \end{subfigure}
    \hfill
    \begin{subfigure}{0.22\linewidth}
    \centering
        \includegraphics[width=\linewidth]{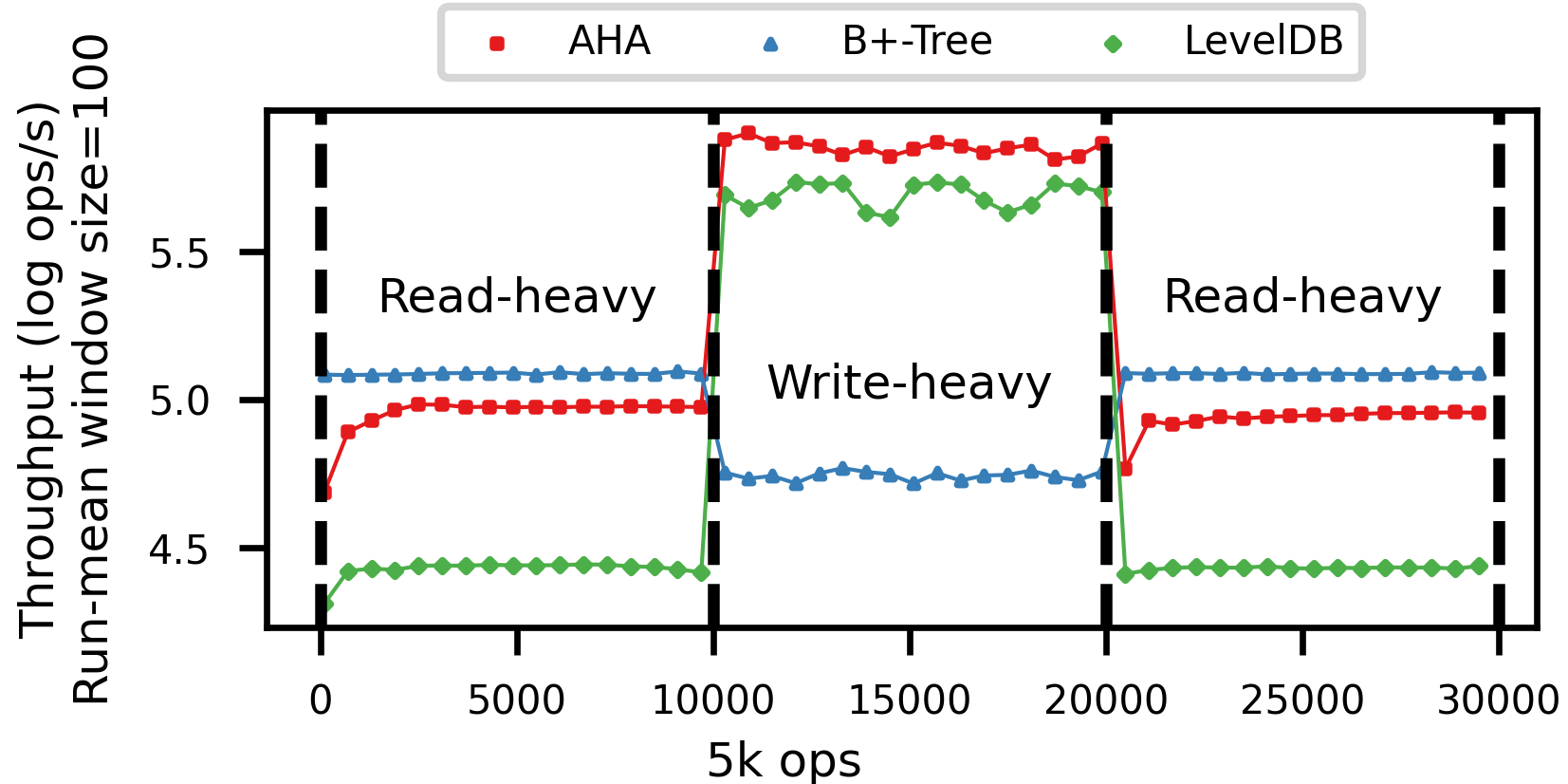}
        \caption{Zipfian dataset with oscillating workloads.}
        \label{fig:zipf_baseline}
    \end{subfigure}
    \hfill
    \begin{subfigure}{0.22\linewidth}
    \centering
        \includegraphics[width=\linewidth]{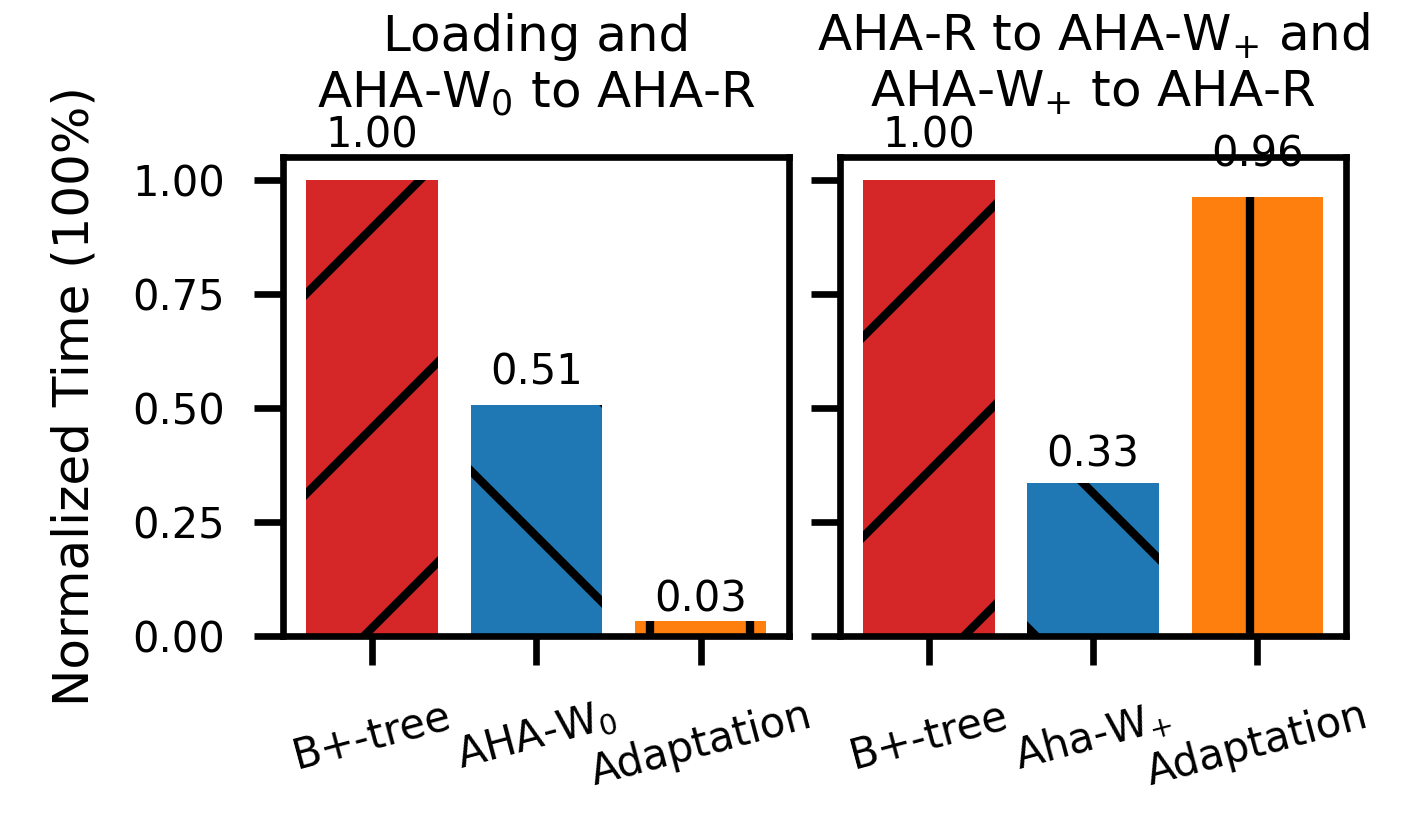}
        \caption{The elapsed adaptation time (Zipfian dataset).}
        \label{fig:zipf_construct}
    \end{subfigure}
    \vfill
    \begin{subfigure}{0.19\linewidth}
    \centering
        \includegraphics[width=\linewidth]{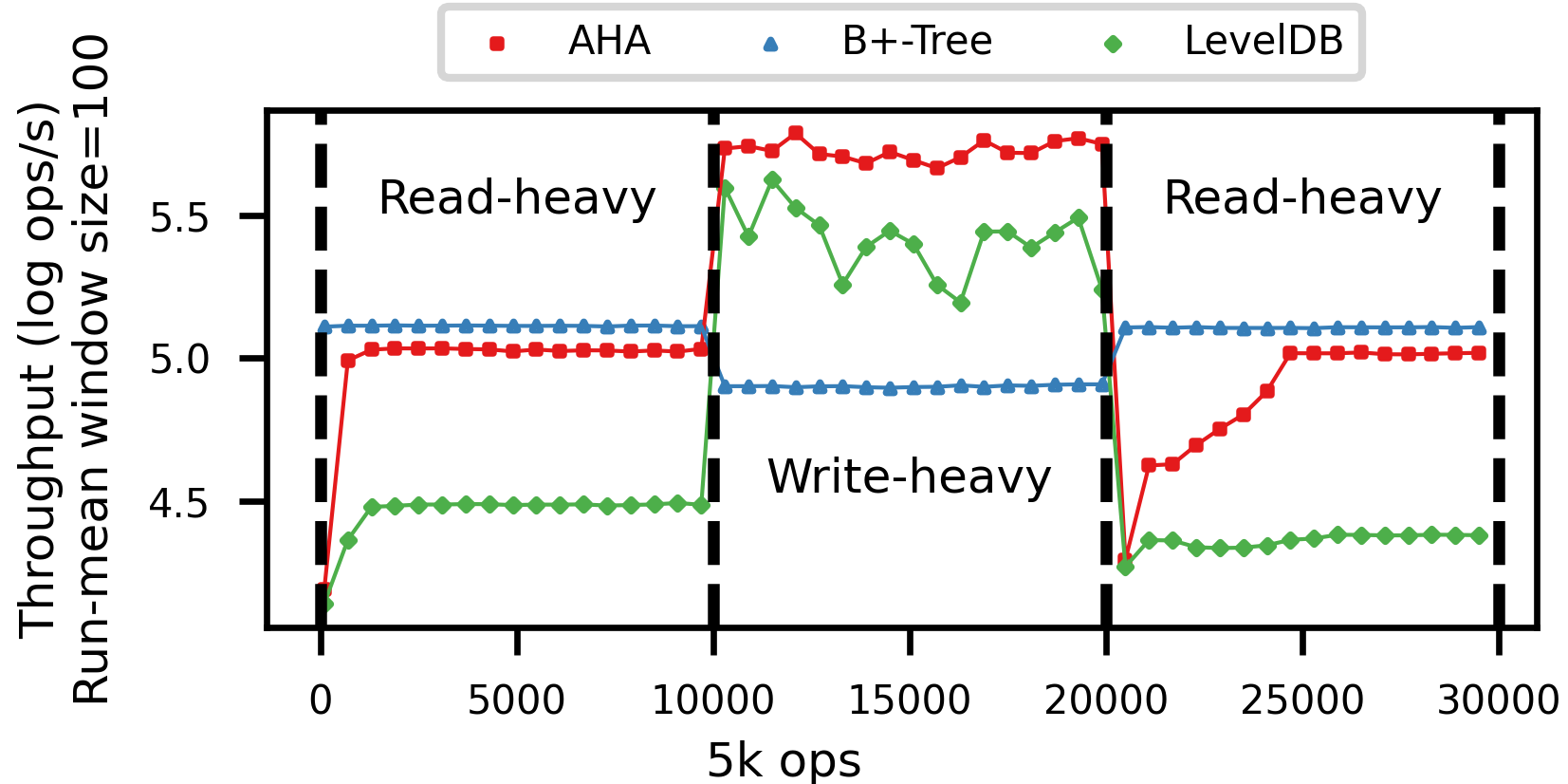}
        \caption{5\% hotspot and range scan selectivity $2\times 10^{-5}$\%.}
        \label{fig:scan100_hot_0.05}
    \end{subfigure}
    \hfill
    \begin{subfigure}{0.19\linewidth}
    \centering
        \includegraphics[width=\linewidth]{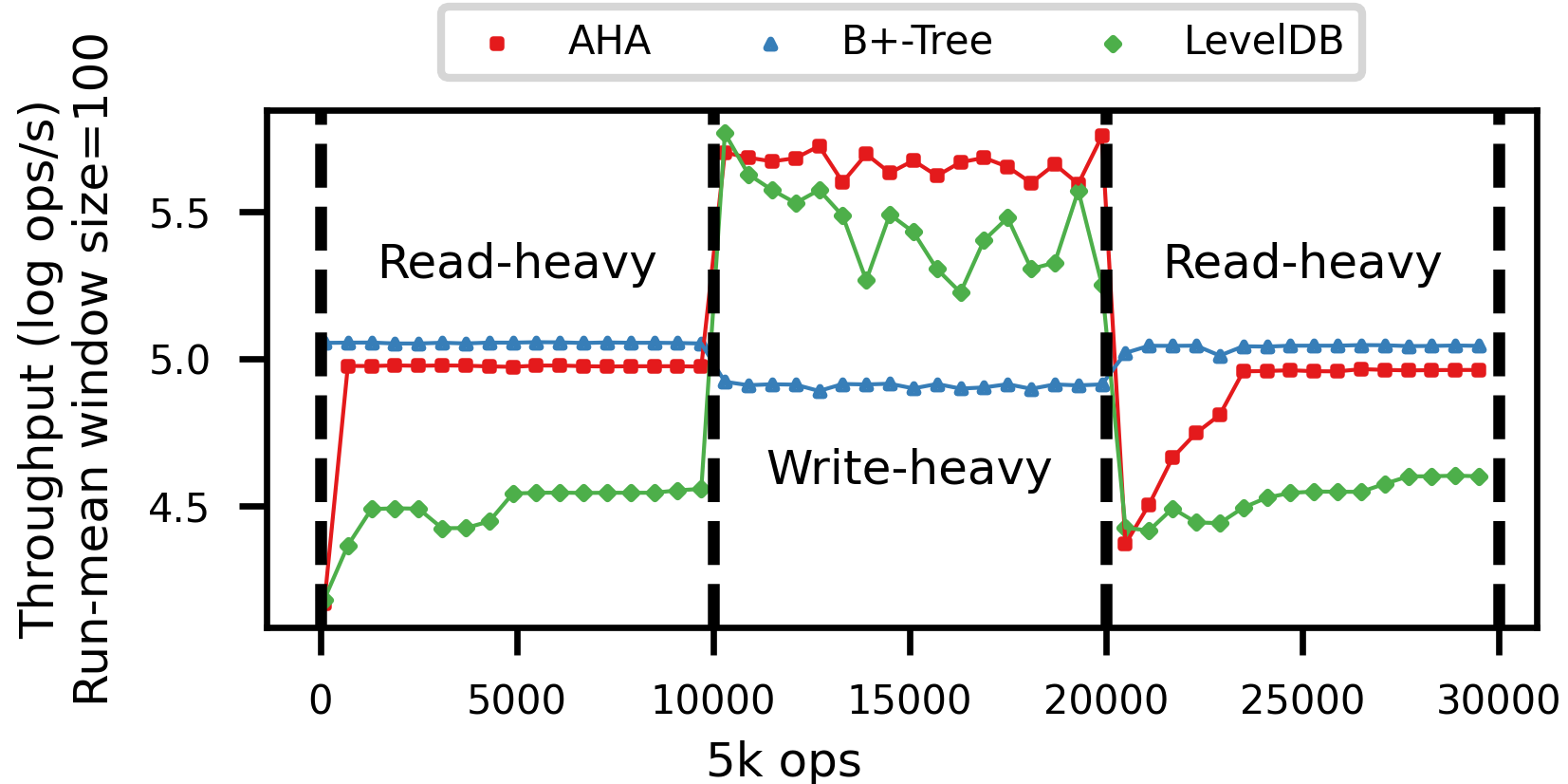}
        \caption{5\% hotspot and range scan selectivity $2\times 10^{-4}$\%.}
        \label{fig:scan1000_hot_0.05}
    \end{subfigure}
    \hfill
    \begin{subfigure}{0.19\linewidth}
    \centering
        \includegraphics[width=\linewidth]{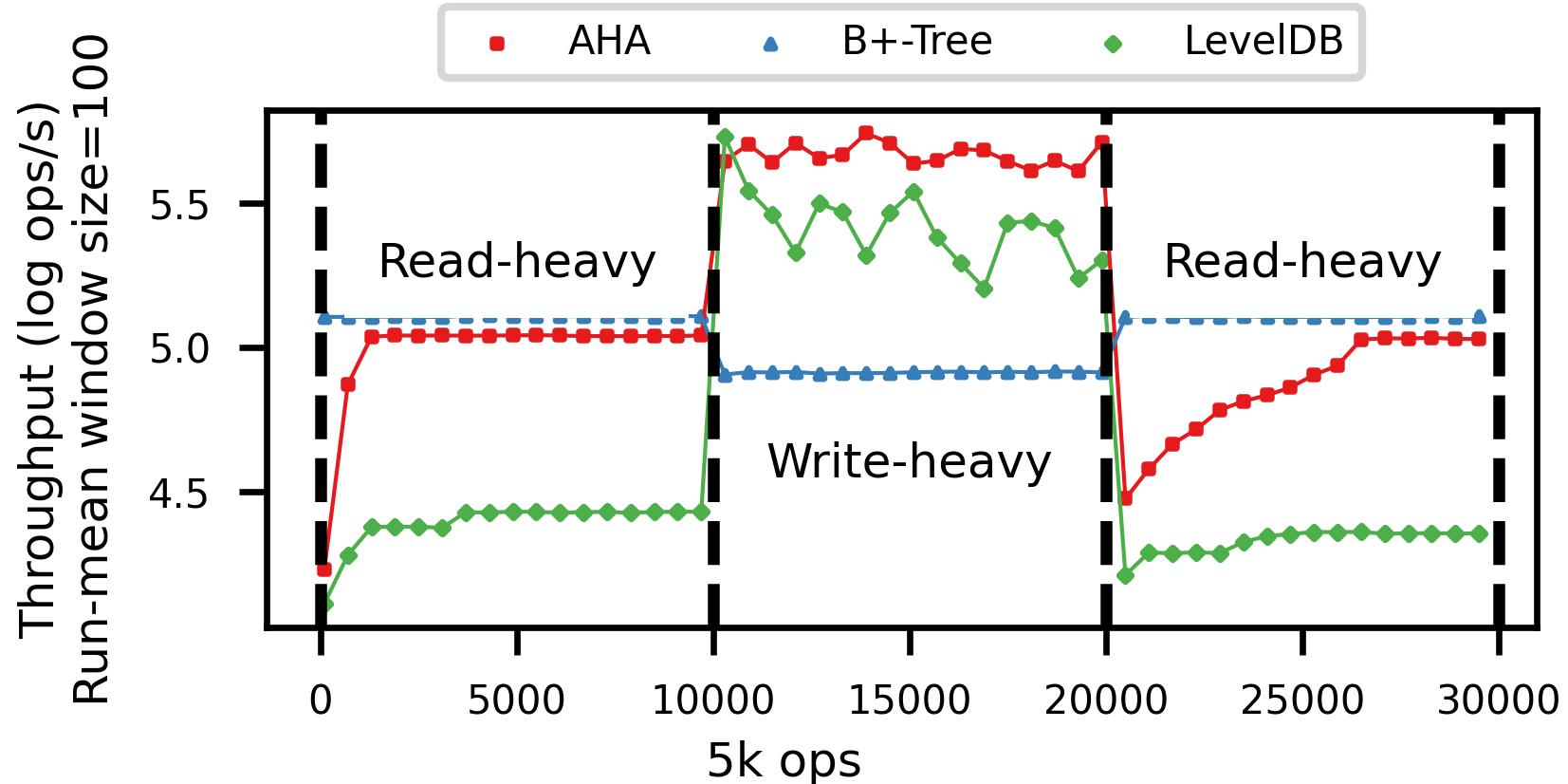}
        \caption{10\% hotspot and range scan selectivity $2\times 10^{-5}$\%.}
        \label{fig:scan100_hot_0.1}
    \end{subfigure}
    \hfill
    \begin{subfigure}{0.19\linewidth}
    \centering
        \includegraphics[width=\linewidth]{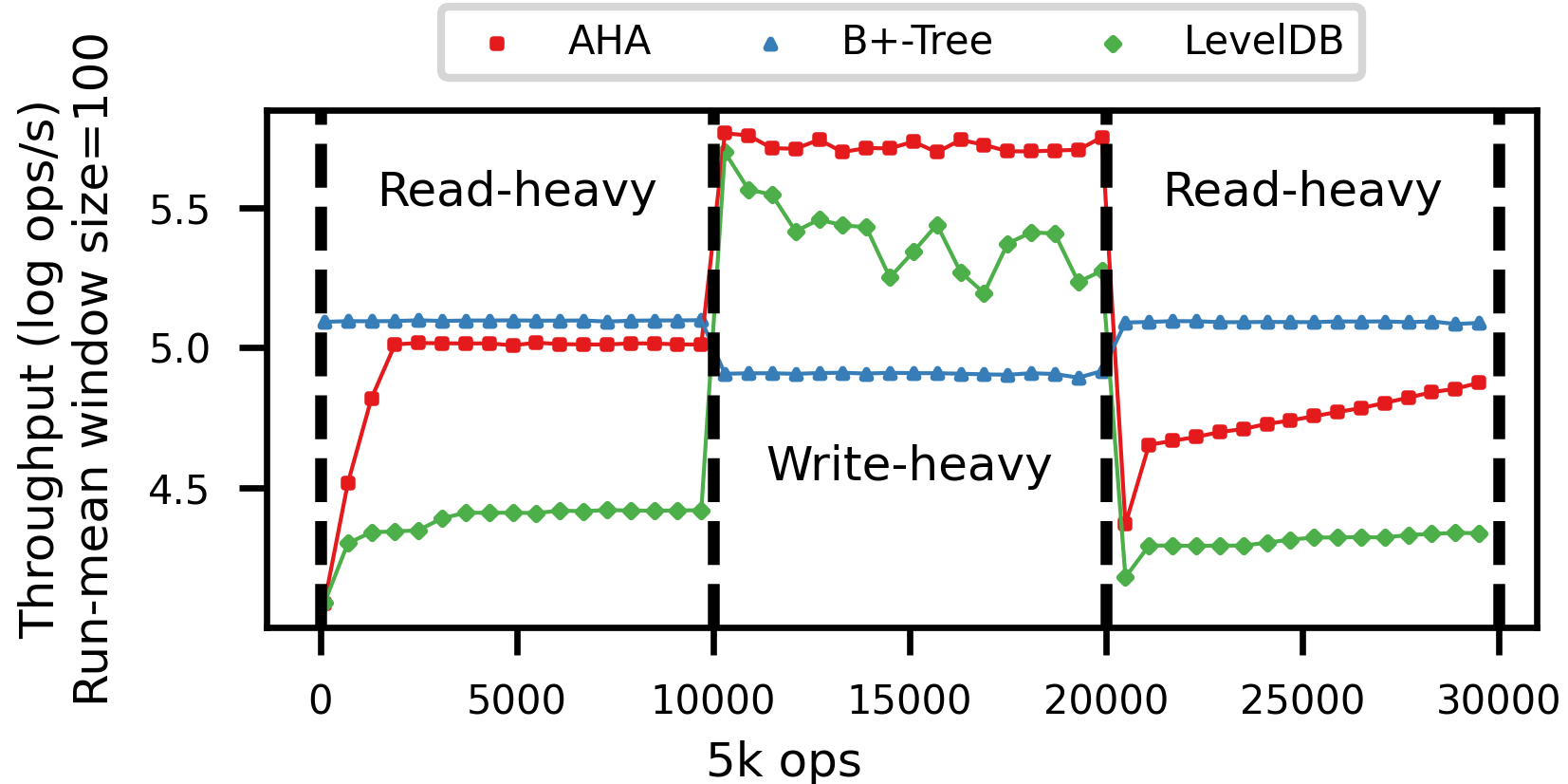}
        \caption{15\% hotspot and range scan selectivity $2\times 10^{-5}$\%.}
        \label{fig:scan100_hot_0.15}
    \end{subfigure}
    \hfill
    \begin{subfigure}{0.19\linewidth}
    \centering
        \includegraphics[width=\linewidth]{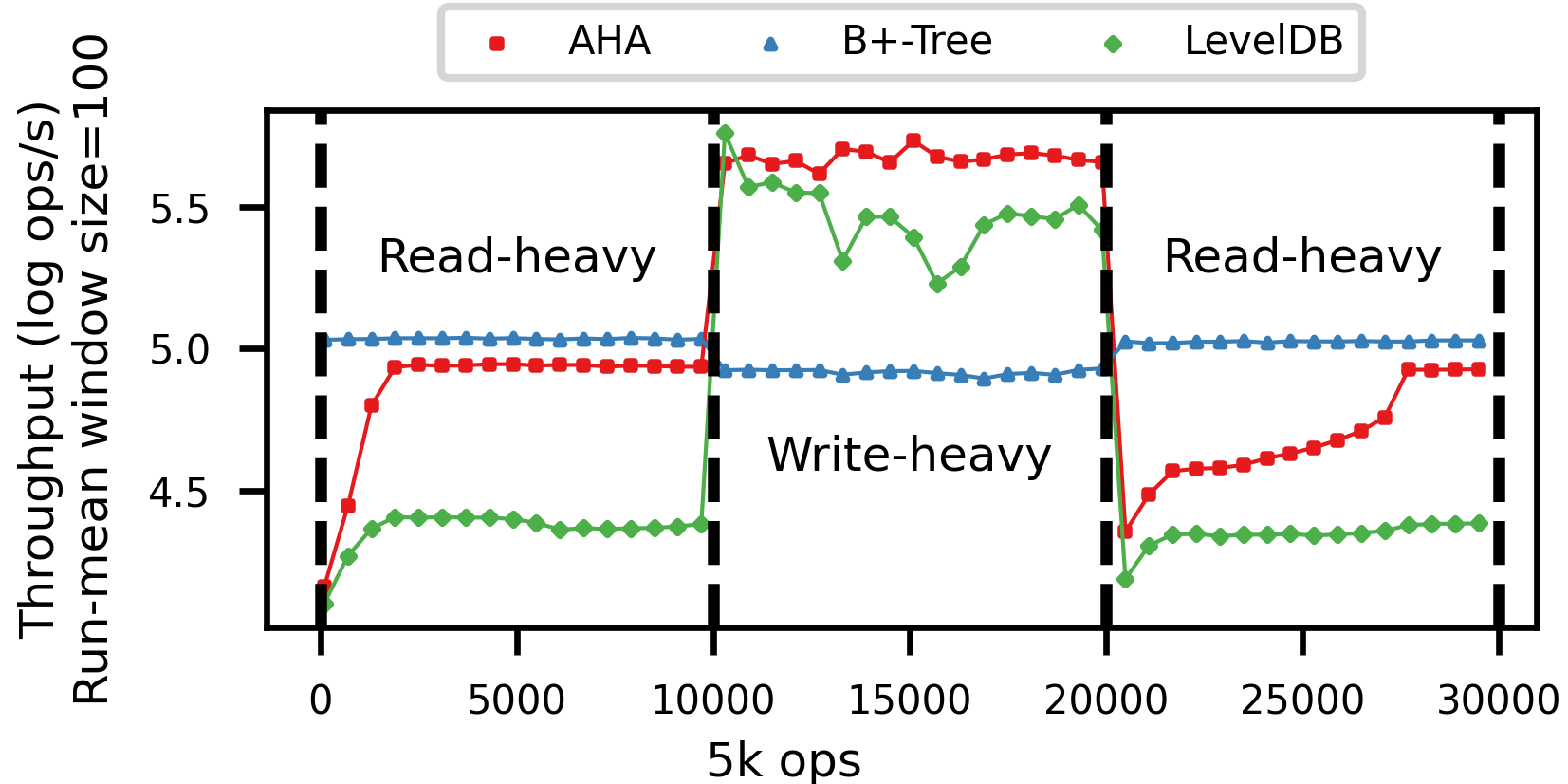}
        \caption{15\% hotspot and range scan selectivity $2\times 10^{-4}$\%.}
        \label{fig:scan1000_hot_0.15}
    \end{subfigure}
    \caption{Performance of the \sys{} under oscillating write- and read-heavy workloads}
    \label{fig:initial_adapt}
\end{figure*}

For our experiments, we use both uniform and Zipfian datasets, and test all indexes using them. The result of the uniform dataset is given in Figures~\ref{fig:unif_baseline} and \ref{fig:unif_construct} while the results of the Zipfian dataset are given in Figures~\ref{fig:zipf_baseline} and \ref{fig:zipf_construct}.
We vary the size of the hotspot and the length of the range query under the uniform dataset to show how adaptation can be affected (Figures~\ref{fig:scan100_hot_0.05}-\ref{fig:scan1000_hot_0.15}).

\subsubsection{\sysw{} to \sysr{} Adaptation}
In Figure~\ref{fig:unif_baseline}'s first 50 Million operations (range queries-only), we observe that for the \sys{}, the throughput gradually increases and reaches \bplustree{} after 1.5k operations. The throughput is 3.24$\times$ that of LevelDB. We also measure the elapsed time. The index construction time of the \bplustree{} is the baseline. The index construction time of \sysw{} is 0.58$\times$ of the \bplustree{} time (Figure~\ref{fig:unif_construct}) and the adaptation time from \sysw{} to \sysr{} is 0.04$\times$ of the \bplustree{} construction time.
For the Zipfian dataset, index construction of \sysw{} takes 0.51$\times$ and the adaptation takes 0.03$\times$ of the \bplustree{} index construction time as in Figures~\ref{fig:zipf_baseline} and~\ref{fig:zipf_construct}. The throughput of \sysr{} is 3.58$\times$ that of LevelDB.

We vary the size of the hotspot as well as the selectivity of the range query. Comparing Figures~\ref{fig:scan100_hot_0.05}, \ref{fig:scan100_hot_0.1} and \ref{fig:scan100_hot_0.15},  notice that, with larger hotspot size, \sysw{} takes more operations to be completely adapted to \sysr{}. This is due to the larger number of data items that need to be adapted. If we compare the same hotspot size but different range query selectivities (cf. Figures~\ref{fig:scan100_hot_0.05} and \ref{fig:scan1000_hot_0.05}), the \sysw{} to \sysr{} adaptation takes roughly the same amount of operations to finish.

\subsubsection{\sysr{} to \sysw{} Adaptation}
In Figure~\ref{fig:unif_baseline} from Operation 50 Million to 100 Million (the write-heavy phase), \sysr{} transitions to \syswp{} that inserts faster than \bplustree{} inserts, and is more stable than LevelDB. The reason is that LevelDB blocks the write operations to process  background compaction. However, in \syswp{}, although it has one fewer insertion thread (to guarantee the total number of threads being the same), a faster and nonblocking compaction is guaranteed by the flexible size limit and smaller compacted \sst{}s because of the guarded compaction (ref Section~\ref{section:implementation}). 
\syswp{} takes 0.64$\times$ of the time needed by the \bplustree{} to finish writing the same amount of data (Figure~\ref{fig:unif_construct}).
For Zipfian dataset, inserting 50 Million data items into \syswp{} takes 0.33$\times$ time of the \bplustree{'s} write time as  Figure~\ref{fig:zipf_construct} illustrates. 

For varying hotspot sizes, no difference is observed during the write-heavy phase when \sysr{} transitions to \syswp{} (Figures~\ref{fig:scan100_hot_0.05}-\ref{fig:scan1000_hot_0.15}).

\subsubsection{\syswp{} to \sysr{} Adaptation}
In Figure~\ref{fig:unif_baseline}, from Operation 100 Million to 150 Million (i.e., the second read-heavy phase), after 25 Million operations, \syswp{} finishes transitioning to \sysr{}. The elapsed transitioning adaptation time is 1.19$\times$ of the \bplustree{} write time. The throughput of \sysr{} is 2.72$\times$ that of LevelDB.
For the Zipfian dataset, after 50 Million data items are inserted into \syswp{}, 
we compare the transitioning time from \syswp{} to \sysr{} with the time taken by the \bplustree{} to absorb 50 Million writes. Transitioning of \sys{} takes 0.96$\times$ of the \bplustree{} time
as shown in Figure~\ref{fig:zipf_construct}.

With larger hotspot sizes, the adaptation process  takes more operations to finish, as demonstrated in Figures~\ref{fig:scan100_hot_0.05}, \ref{fig:scan100_hot_0.1} and \ref{fig:scan100_hot_0.15}. For Hotspot Size 15\% with Range Query Selectivity $2\times 10^{-5}$\%, the transition 
takes nearly 43 Million operations to finish.
When the range query selectivity is larger, this adaptation process can finish faster, as shown in Figures~\ref{fig:scan100_hot_0.15} and \ref{fig:scan1000_hot_0.15}. \syswp{} in Figure~\ref{fig:scan1000_hot_0.15} can complete transitioning to \sysr{} within 50 Million operations. The reason is that with larger range query length, more \ndl{}s are visited by the range queries, and these \ndl{}s are queued for further adaptation.
With more nodes being adapted per range query, this speeds up the adaptation.

After \sysr{} is completely restored, the range query throughput can catch up with the \bplustree{}, and is approximately 4.2$\times$ that of LevelDB. For the Zipfian data, there is a small gap in throughput between \sysr{} and \bplustree{}. No difference is observed with varying hotspot sizes and range query lengths.

\subsubsection{The \sys{} vs. the LSM-Tree}
In LevelDB~\cite{leveldb}, the \sst{} compaction optimization can be triggered by excessive \sst{} reads. It helps compact frequently-read \sst{}s to reduce I/O. An \sst{} is allowed to be seeked 100 times by default. If this number is exceeded, compaction is triggered to compact all the \sst{}s of the overlapping range, and produce new \sst{}s. We observe the effect of this optimization in LevelDB but still cannot catch up with the \bplustree{} nor \sysr{} as shown in Figures~\ref{fig:unif_baseline} and~\ref{fig:zipf_baseline}. 

\subsection{Versatility of Hotspots}
In this section, we explore the versatility of hotpots. Instead of having a fixed single hotspot, real datasets often present multiple hotspots and hotspot may even change their location. We run our experiments using uniform data that includes multiple hotspots (Section~\ref{sssection:multi-hot}) and dynamically changes hotspot locations (Section~\ref{sssection:change-hot}) while the indexes are executing the workload.

\begin{figure}[h]
    \centering
    \begin{subfigure}{0.49\linewidth}
    \centering
        \includegraphics[width=\linewidth]{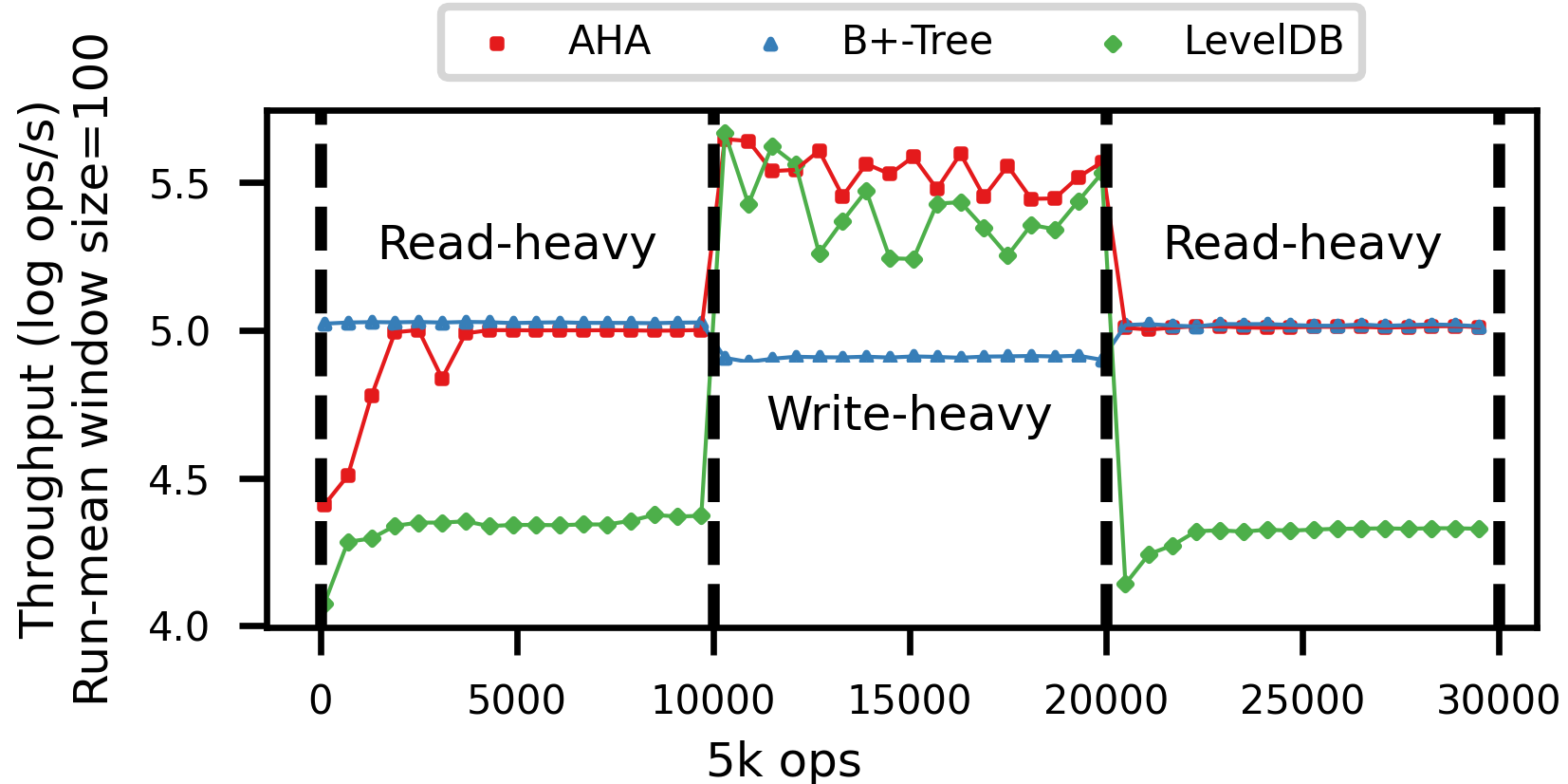}
        \caption{Index throughput with two hotspots.}
        \label{fig:multi_hotspots}
    \end{subfigure}
    \hfill
    \begin{subfigure}{0.49\linewidth}
    \centering
        \includegraphics[width=\linewidth]{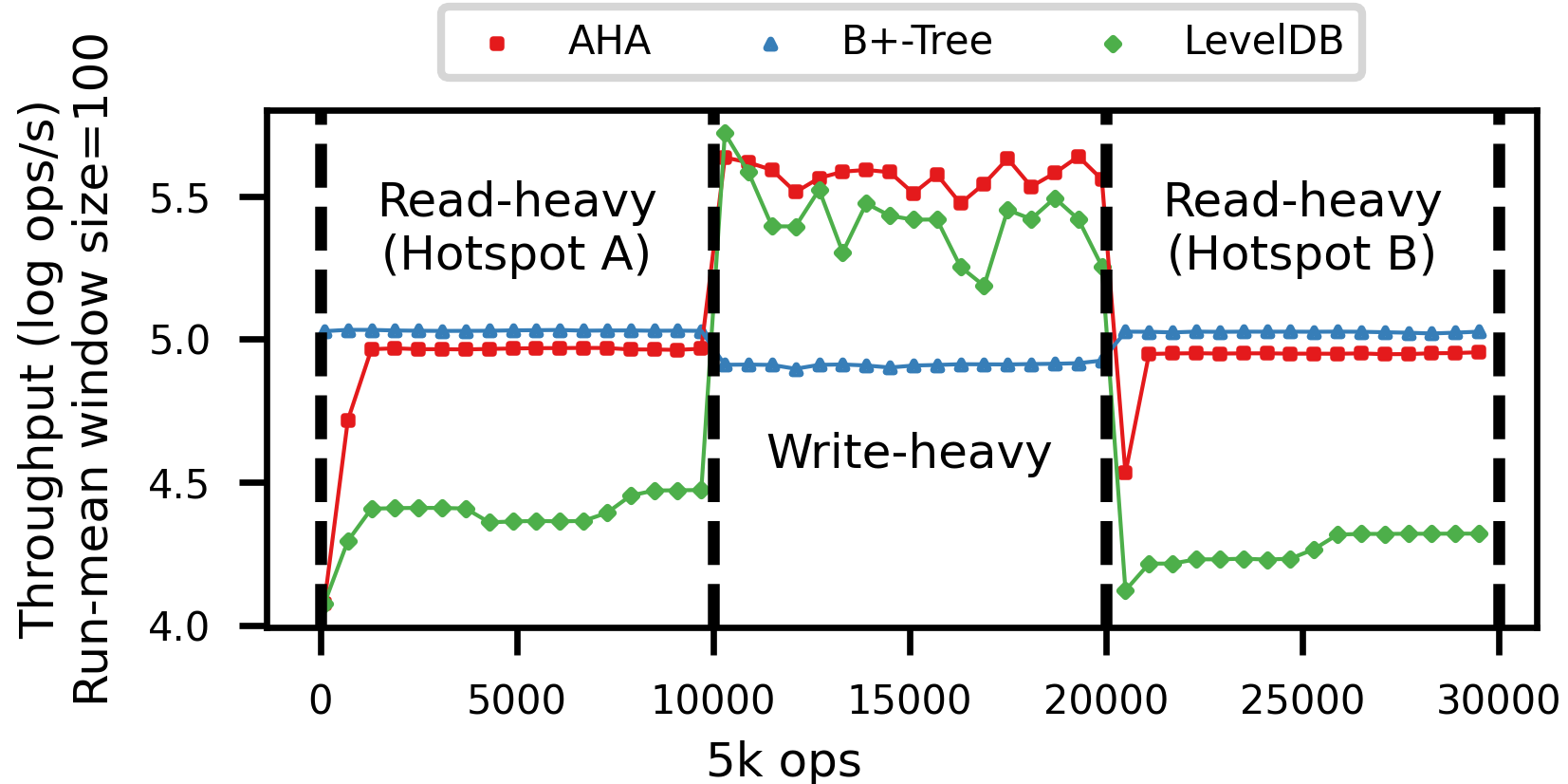}
        \caption{Index throughput (hotspot changes during the 2nd read phase)}
        \label{fig:change_hotspots}
    \end{subfigure}
    \caption{Index performance with versatile hotspots.}
    \label{fig:versatile-hotspot}
\end{figure}

\subsubsection{Multiple Hotspots}\label{sssection:multi-hot}
Real-world datasets often contain multiple hotspots. The \sys{} can adapt itself even under multiple disjoint hotspots. We conduct experiments with two hotspots (Figure~\ref{fig:multi_hotspots}) and demonstrate the feasibility of \sysr{} under multiple hotspots. In the experiment, we add one more hotspot in the middle of the key space besides the original one at the leftmost of the key space. \sysw{} can still transition to \sysr{} during the read phase from 0 to 50 Million operations. During the 2nd read phase from 100 Million to 150 Million, \syswp{} can also transition to \sysr{}. The performance after the transient transitioning time is as good as the \bplustree{} and 4.22$\times$ that of LevelDB.

\subsubsection{Hotspot Drift}\label{sssection:change-hot}
There are cases when hotspots  change their locations, e.g., popular products may change overtime, popular tourism sites may change, etc. The \sys{} still performs well under this scenario (Figure~\ref{fig:change_hotspots}). The initial hotspot for the first 50 Million range-query operations lie in the leftmost of the key space. After the next 50 Million write-only operations, the hotpot is changed to the middle of the key space and remains there until the end. We observe for both  range-query-only phases (0-50 Million and 100-150 Million operations), both \sysw{} and \syswp{} can adapt to \sysr{} completely that is comparable with the \bplustree{} in read performance.

The versatility of hotspots makes the \sys{} more appealing. 
With a single static hotspot, 
it is attempting to use two separate indexes, e.g., the \bplustree{} and the \lsmt{} to store hot data and cold data, respectively. 
However, with more hotspots or hotspots that keep changing locations, we need to keep track of the index that stores the freshest data item for each key-value pair. An item \texttt{D} that is initially in the hotspot should reside in the \bplustree{}. If \texttt{D} becomes cold, it should be moved to the \lsmt{}. But when \texttt{D} becomes hot again, it should be in the \bplustree{} to facillitate fast search.  It is a nontrivial task to migrate a large number of data items between indexes back and forth as hotspots keep moving.

\begin{figure}[h]
    \centering
    \begin{subfigure}{0.49\linewidth}
    \centering
        \includegraphics[width=\linewidth]{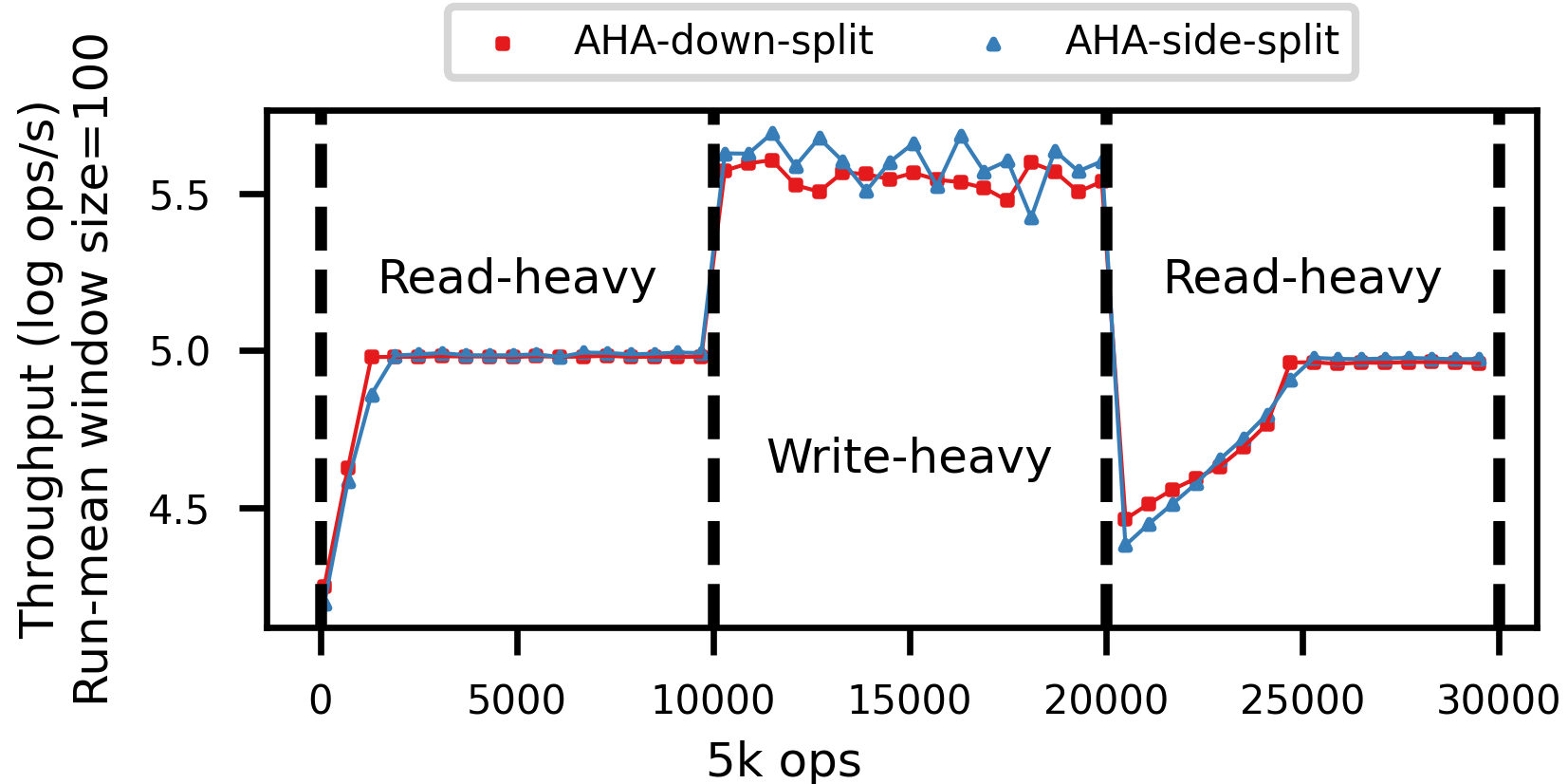}
        \caption{Performance overtime with uniform dataset}
        \label{fig:down-side-unif}
    \end{subfigure}
    \hfill
    \begin{subfigure}{0.49\linewidth}
    \centering
        \includegraphics[width=\linewidth]{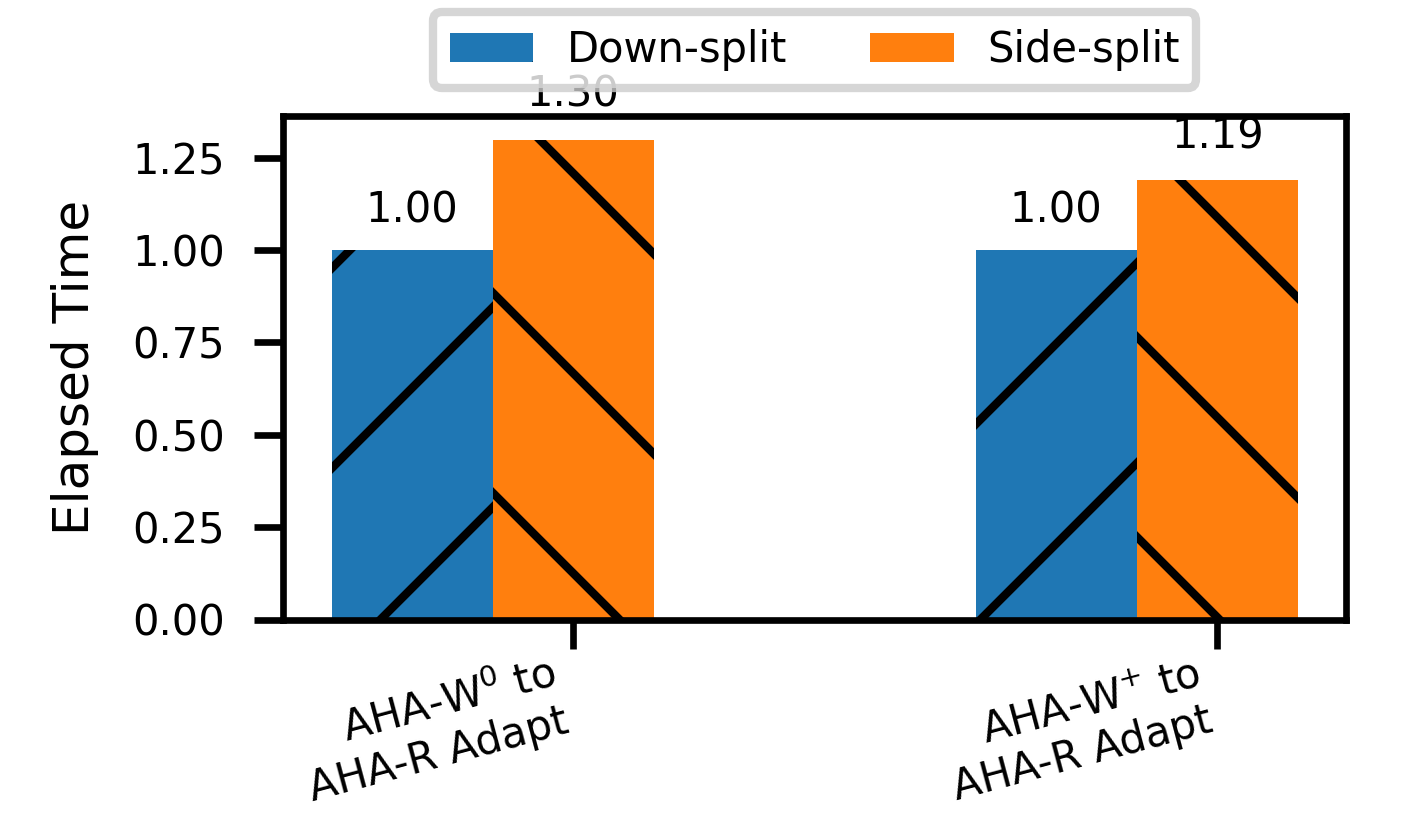}
        \caption{Time comparison with uniform data}
        \label{fig:down-side-unif-time}
    \end{subfigure}
    \vfill
    \begin{subfigure}{0.49\linewidth}
    \centering
        \includegraphics[width=\linewidth]{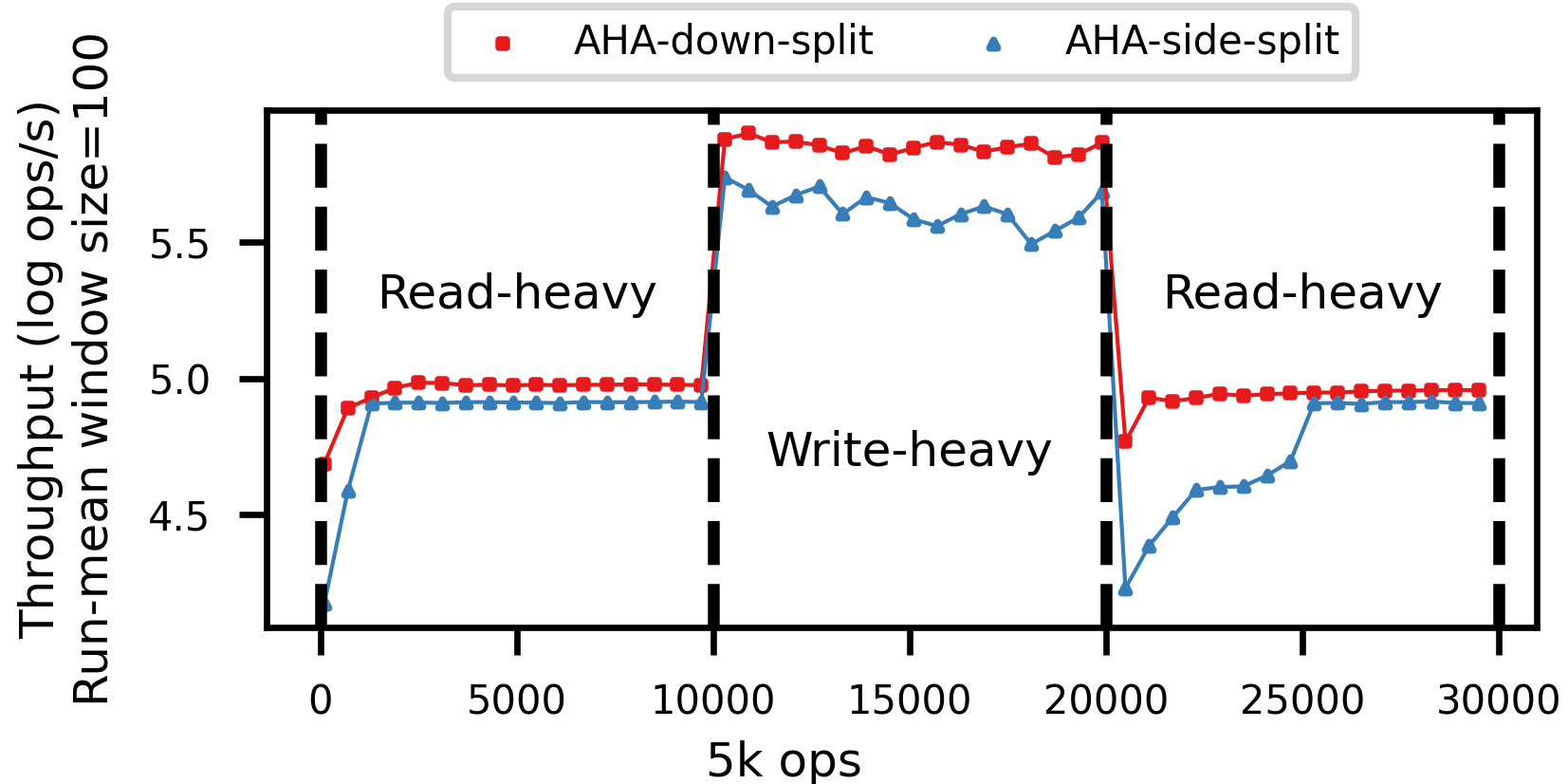}
        \caption{Performance overtime with Zipfian dataset}
        \label{fig:down-side-zipf}
    \end{subfigure}
    \hfill
    \begin{subfigure}{0.49\linewidth}
    \centering
        \includegraphics[width=\linewidth]{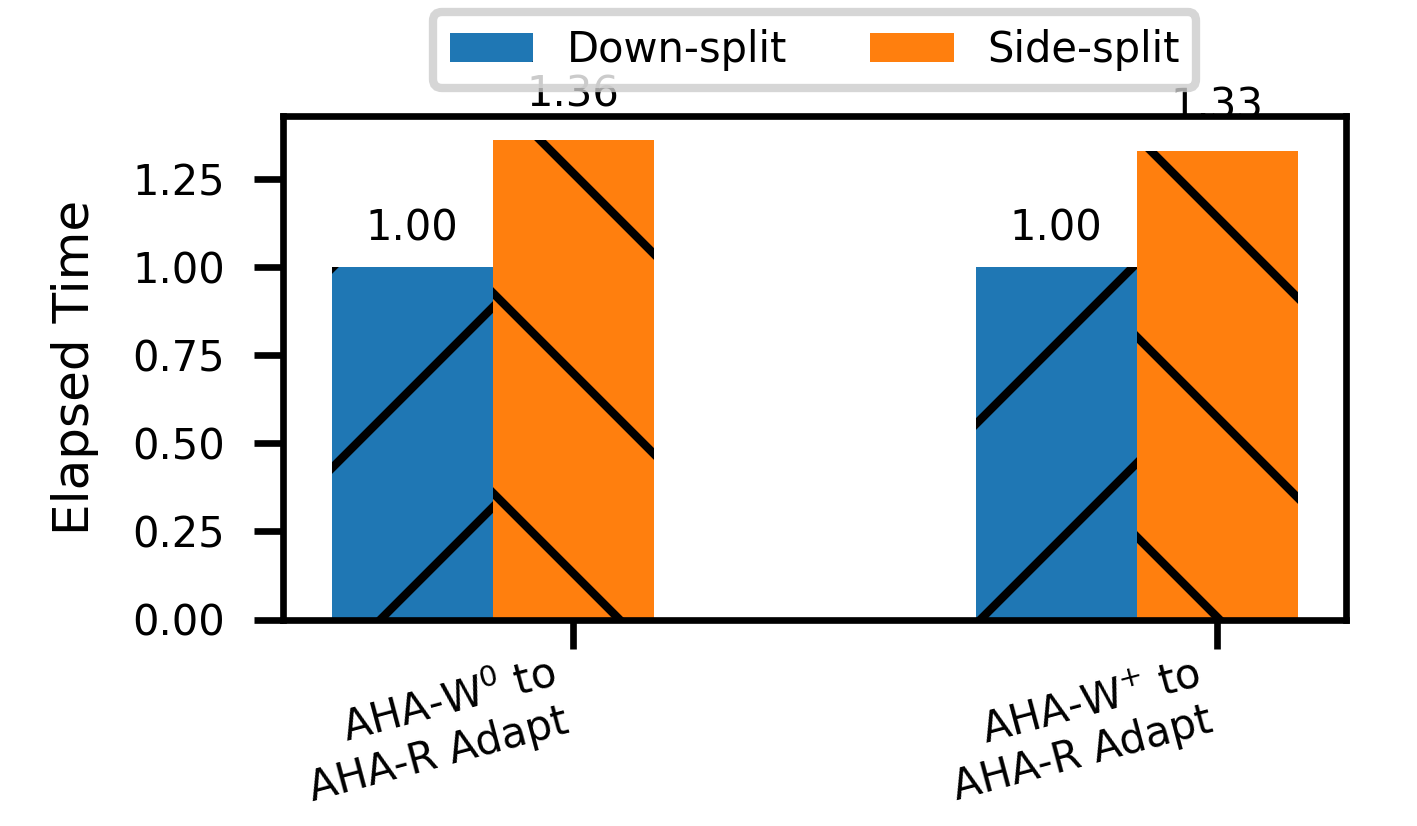}
        \caption{Time comparison with Zipfian data}
        \label{fig:down-side-zipf-time}
    \end{subfigure}
    \caption{Comparison of two leaf adapting techniques down-split and side-split under the uniform and Zipfian datasets.}
    \label{fig:down-side}
\end{figure}

\subsection{Efficiency of Adaptation}
We have proposed down-split 
and side-split
in Section~\ref{sssection:w2r}. We compare the two splitting mechanisms using the uniform and Zipfian datasets in Figure~\ref{fig:down-side}. Overall, both techniques can adapt \sysw{} to \sysr{}, and \syswp{} to \sysr{}. Figures~\ref{fig:down-side-unif-time} and \ref{fig:down-side-zipf-time} present the relative time taken to adapt. For both datasets, side-split takes longer time to be fully adapted to \sysr{}. For the uniform dataset, side-split takes 1.30$\times$ time to adapt \sysw{} and 1.19$\times$ time to adapt \syswp{}. For the Zipfian dataset, side-split takes 1.30$\times$ time to adapt \sysw{} and 1.33$\times$ time to adapt \syswp{}. This is expected as side-split compacts the \sst{s} in the leaf \ndl{} to assign them to different leaf pages, while down-split adds a new level in the middle and saves the effort of compacting the \sst{}s. 

\subsection{Analysis of Adaptation}
We explore the factors that contribute to the \sys{} adaptation time. 

\subsubsection{Effect of \lsmt{} Size}
In previous experiments, we set the limit of the number of levels for \rtl{} to 3,  and \ndl{} to 2. We vary the limit for both \rtl{} and \ndl{} and compare the adaptation process in Table~\ref{tab:lsm-size}.



\begin{table}[h]
    \centering
    \begin{tabular}{c|c|c|c|c|c}
    (Root, Node) & (3, 2) & (4, 2) & (2, 2) & (3, 1) & (3, 3)\\
    \hline
     \sysw{} to \sysr{} Adapt (T) & 1 & 0.66 & 1.04 & 4.58 & 2.82\\
     \sysw{} to \sysr{} Adapt (O) & 1 & 1.01 & 1.04 & 4.01 & 2.54\\
     \syswp{} to \sysr{} Adapt (T) & 1 & 1.01 & 1.20 & 1.59 & 0.72\\
     \syswp{} to \sysr{} Adapt (O) & 1 & 1.02 & 1.19 & 1.67 & 0.69\\
    \end{tabular}
    \caption{Effect of \lsmt{} size measured in relative elapsed time (T) and number of operations (O). (Root, Node) corresponds to the pairs of level limit of \rtl{} and \ndl{}, e.g., (3, 2) represents \rtl{} that has 3 levels and \ndl{} that has 2 levels.}
    \label{tab:lsm-size}
\end{table}

For the \sysw{} to \sysr{} adaptation time, the size of \rtl{} does not have significant effect as pairs (3, 2), (4, 2) and (2, 2) do not differ much in the first two rows. However, changing the size of \ndl{} to either one or three both increase the initial adapt time (cf. columns (3, 1) and (3, 3) first two rows). When \ndl{} has only one level, this structure resembles the buffer tree~\cite{arge2003buffer}. By default, the single-leveled \lsmt{} has the \sst{}s overlapping  each other in Level-0. During the node-emptying process, since these \sst{}s have not undergone the guarded compaction, the \sst{} ranges are not aligned with the pivot keys in the tree page. Thus,  adaptation takes longer time. However, when there are 3 levels in \ndl{}, each \ndl{} now has a larger capacity to hold more data items. Thus, more hot data exists in \ndl{}, and this results in a longer adaptation time.


For the \syswp{} to \sysr{} transition time, the size of \rtl{} does not have significant effect as well (cf. columns (3, 2), (4, 2) and (2, 2) bottom two rows). A single-leveled \ndl{} still takes longer time to adapt. However, a three-leveled \ndl{} requires lesser time to adapt. A (3, 3) configuration
results in an \sys{} with smaller height than a (3, 2) configuration.
And during the write-heavy phase from Operation 50 Million to 100 Million, the (3, 3) configuration takes 1.84$\times$ more time than the (3, 2) one, suggesting that a lot of the hot data has already been merged into leaf pages.
Since more hot data has been merged with tree pages already, the 
amount 
of data that accumulates and that needs to be adapted becomes fewer with the lesser time.

\subsubsection{Effect of Multi-threading}
We vary the number of user threads that perform operations on the index. 


\begin{table}[h]
    \centering
    \begin{tabular}{c|c|c|c|c}
         Number of Threads & 3 & 6 & 10 & 16\\
         \hline
         \sysw{} to \sysr{} Adapt (T) & 1 & 0.65 & 1.06 & 0.89\\
         \sysw{} to \sysr{} Adapt (O) & 1 & 1.40 & 3.31 & 3.07\\
         \sysr{} to \syswp{} and Finish (T) & 1 & 0.87 & 0.85 & 0.83\\
         \syswp{} to \sysr{} Adapt (T) & 1 & 1.11 & 1.02 & 1.29\\
         \syswp{} to \sysr{} Adapt (O) & 1 & 1.89 & 2.59 & 2.96\\
    \end{tabular}
    \caption{Effect of Multi-threading}
    \label{tab:my_label}
\end{table}

From the table, with 6 threads, the \sysw{} to \sysr{} adaptation time is the smallest among all. With less threads (3 threads) or more threads (10, 16 threads), this
adaptation time is larger. 
With many threads performing range queries, the threads can visit the same set of \ndl{}s at the same time.
For example, Node A is visited by one user thread. While \texttt{BGtree} is in the progress of adapting Node A, Node A may be added to the work queue by another user thread, causing \texttt{BGtree} to perform some redundant checking of Node A. 
The same is true for the case of fewer threads
where only a limited number of \ndl{}s are visited. Under these two situations, the background threads may not work at 
their
full capacity. During the write phase (50-100 Million operations), 
it takes lesser time to finish with more threads. For the \syswp{} to \sysr{} adaptation, 
having 
too many threads (more than 16 threads) does not improve on the time with the 
same
reason of \sysw{} to \sysr{} adaptation.

\subsubsection{Time Breakdown}
\begin{figure}[h]
    \centering
    \includegraphics[width=0.8\linewidth]{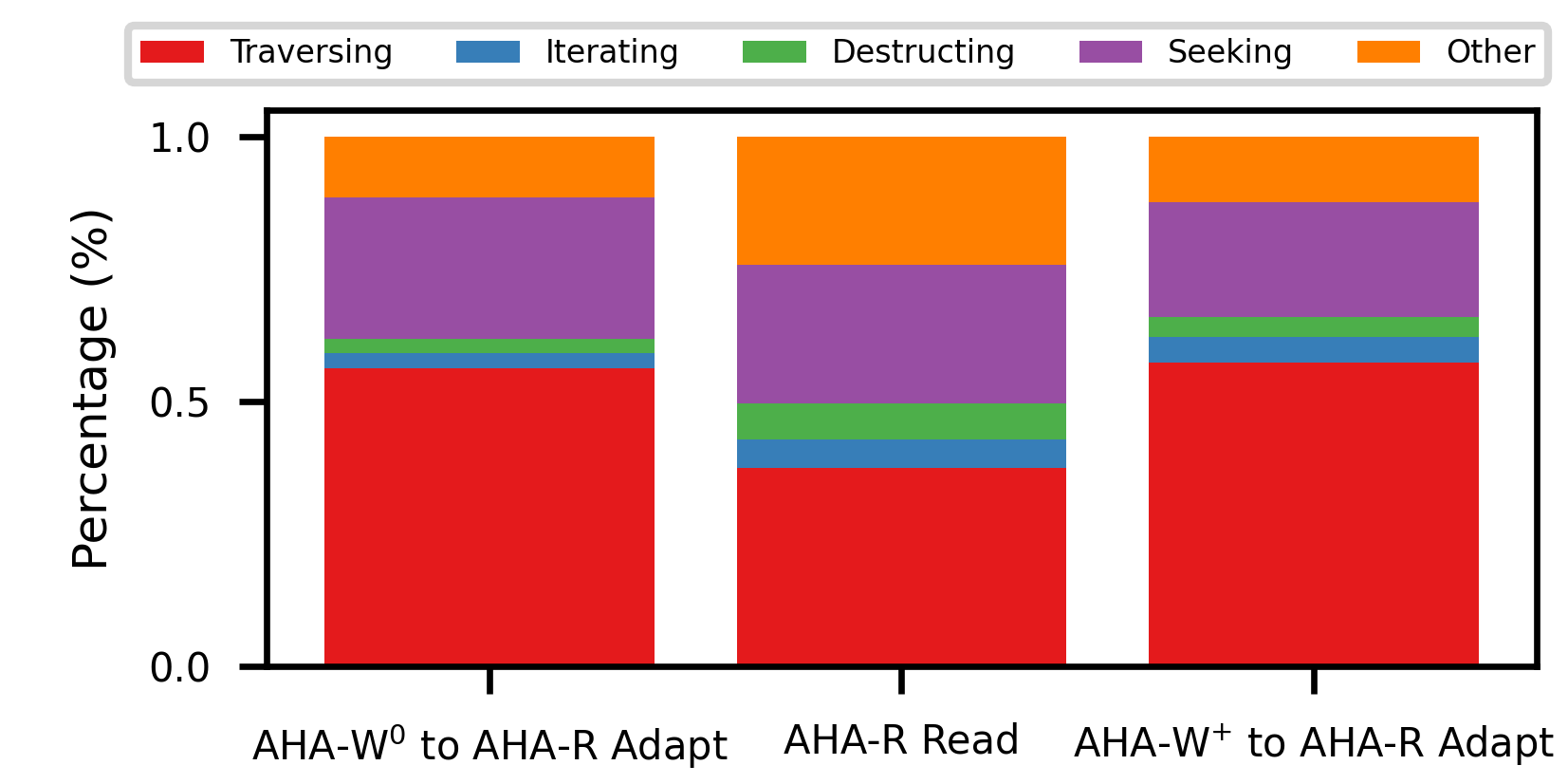}
    \caption{Read time breakdown}
    \label{fig:breakdown}
\end{figure}

We analyze the time breakdown during read-only phases (i.e., during the Operations 0 to 50 Million and  100 to 150 Million). When adaptation is in progress, the majority of the time is on traversing the tree structure to find the target tree node or page. The reason is that when reading and adaptation happen concurrently, reader threads spend more time waiting to obtain the lock as the background thread is adapting.
One possible solution is to apply a multi-word compare-and-swap operation as in the case of the Bztree~\cite{arulraj1028bztree}.

\subsubsection{Write Amplification}
Write amplification is a known issue for the \lsmt{} structure. We compare the write amplification of LevelDB and the \sys{}. The size of the dataset we use is 69 GB. The total write I/O of LevelDB is 1244 GB, which is an 18.02$\times$ write amplification. Constructing a \sys{} takes 821 GB write I/O, which is 12.07$\times$ write amplification. The smaller write amplification is due to the guarded compaction of \rtl{} and \ndl{}s. Also, if we combine \rtl{} and \ndl{}s and view them as one large \lsmt{}, the tree structure with pivoting keys of the \sys{} serve as guards themselves, which further reduces \sst{} rewriting and write amplification. A similar  effect of has observed in PebblesDB~\cite{raju2017pebblesdb}.

\subsubsection{Potential Optimization}\label{sssection:potential_opt}
\begin{figure}[h]
    \centering
    \includegraphics[width=0.5\linewidth]{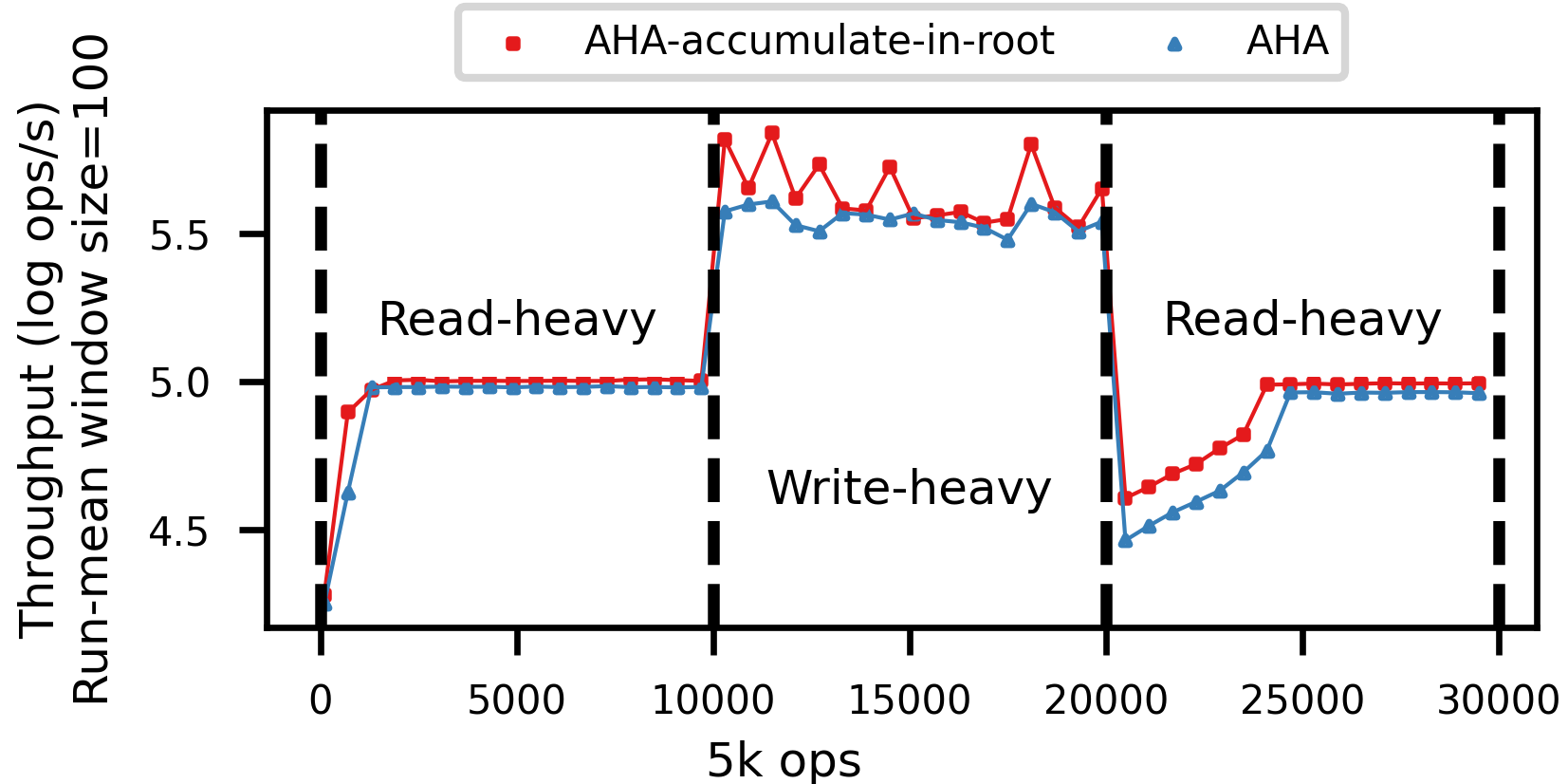}
    \caption{Potential optimization from \syswp{} to \sysr{}}
    \label{fig:potential}
\end{figure}
To speed up the transitioning from \syswp{} to \sysr{},
we may allow \rtl{} to hold all hotspot data during the write-heavy phase (50-100 Million operations). The improvement is demonstrated in Figure~\ref{fig:potential} where the transition 
is faster in contrast to the \sys{} without this optimization.
A comparison with baselines is presented in Figure~\ref{fig:teaser}, showing that the relative time for the \syswp{} to \sysr{} transition is 0.65$\times$ the time required by a \bplustree{} to complete the write-heavy phase. We plan to investigate this and other potential optimizations to enhance the performance further.

\section{Conclusion}\label{section:conclusion}
In this paper, we present a kind of workload that oscillates between write-heavy and read-heavy. This workload is observed in many real-world applications. Traditional non-adaptive indexing and adaptive indexing cannot fit into this workload well as they are either non-adaptive or adapt in only one direction. With the observation that real data sets are skewed, we focus on the hotspots. We encapsulate the adaptation techniques in the \sys{} to make it adaptive in both directions. In the experiments, we show that \sys{} can efficiently adapt itself under oscillating write-heavy and ready-heavy workloads.

\section{Acknowledgements}
L. X. thanks Maxime Schoemans and Ahmed Mahmood for their discussions during the initial stages of this work.
Walid G. Aref acknowledges the support of the National Science Foundation under Grant Number IIS-1910216.

\bibliographystyle{acm}
\bibliography{references}  






\end{document}